\documentclass[aps,prb,twocolumn,preprint,10pt]{revtex4-1}
\usepackage{overpic}
\usepackage{amsmath,amssymb}
\usepackage{graphicx,graphics}
\usepackage{bm}
\usepackage{ulem}
\usepackage{hyperref}
\usepackage{physics}
\usepackage{tikz}
\usepackage[inline]{enumitem}
\usepackage{siunitx}
\usepackage{chemformula}

\graphicspath{{figures/}}

\begin{document}

\author{Thomas Benjamin Smith}
\email{tommy.smith023@gmail.com}
\address{Department of Physics and Astronomy, School of Natural Sciences, Faculty of Science and Engineering, University of Manchester, Oxford Road, Manchester, M13 9PY, United Kingdom.}
\author{Coskun Kocabas}
\address{The National Graphene Institute, University of Manchester, Booth Street East, Manchester, M13 9PL, UK}
\author{Alessandro Principi}
\address{Department of Physics and Astronomy, School of Natural Sciences, Faculty of Science and Engineering, University of Manchester, Oxford Road, Manchester, M13 9PY, United Kingdom.}

\title{Topological Surface-plasmon-polaritons on Corrugated Metal-dielectric Surfaces}

\begin{abstract}
We study topological surface-plasmon-polaritons at optical frequencies in diffraction gratings formed by bipartite corrugated metal-dielectric gratings. To do so we implement the theory as developed in \citet{DellaValle:2010} to describe the amplitude of the field by an emergent Schr{\"o}dinger-like equation. The tri-harmonic grating generates a bipartite Kronig-Penney model. Topologically protected localised modes are then predicted to occur at the edges of the grating and at defects formed by the combination of two mirror antisymmetric corrugations, whose bulk invariant is a step-wise varying Zak phase in both cases.
\end{abstract}

\maketitle

\section{Introduction}

The topic of 1D topological insulators is an established and well-ploughed field\cite{TKNN,Thouless:1994,Kohmoto:1985,Novoselov:2006,Zhang:2015,Avron:2003,Hatsugai:1997,Gurarie:2013,Aoki:1986,Watson:1996}. Beginning with the study of the one-dimensional Su-Schrieffer-Heeger (SSH) model\cite{SSH:1979,SSH:1980,SSH:1988,Asboth:2016,Kane:2013} as applied to the edge-state properties of electrons in long-chain polymers, the field has since evolved to further low-dimensional (meaning also 2D) systems whose excitations may be, but are not limited to being, photonic\cite{Liu:2018,Wang:2018,Ozawa:2019,Gorlach:2017}, magnonic\cite{Mei:2019,Qin:2017,Pirmoradian:2018}, phononic\cite{Pal:2018,Zhao:2018,Zhang:2018}, acoustic\cite{Yang:2015,Esmann:2018,Jia:2018}, and plasmonic\cite{Poddubny:2014,Downing:2017,Downing:2018,Kruk:2017,Pocock:2018,Yousefi:2019,Wang:2016}.

The standard approach in all of these systems is to assume the tight-binding formalism. Then the constructed low-energy effective Hamiltonian, which generates the Schr{\"o}dinger equation, may possess certain discrete symmetries. When one such symmetry, the chiral one, is present in one-dimension then topologically protected edge modes will be present. The appeal of such modes is both theoretical, for their fundamental interest due to the phenomenon of bulk-boundary correspondence, and experimental, for their robustness against bulk defects and disorder due to the topological invariance of the Hamiltonian through chiral-symmetry-preserving adiabatic deformations.

In the context of topological plasmonics\cite{Poddubny:2014,Downing:2017,Downing:2018,Kruk:2017,Pocock:2018,Yousefi:2019,Wang:2016}, the tight-binding approximation may be made if the lattice in question is built of nanoparticles that are capable of hosting localised-plasmon-polaritons (LPPs). These modes form the basis `atomic wavefunctions' of the tight-binding model and the hopping parameters are simply the tunnelling amplitudes that arise from the dipole interactions of these LPPs between neighbouring nanoparticles. 

Surface-plasmon-polaritons (SPPs), much like LPPs, are quasi-particles formed from the electromagnetic interaction of photons with plasmons. As such, they may only exist at a dielectric-metal interface whereat the sign of the dielectric function changes; a fact which has been recently shown to be of topological origin\cite{Bliokh:2019}. Work has been conducted on the appearance of SPPs at the surface of 3D topological insulators\cite{Qi:2014,Deshko:2016,Stauber:2017,Smith:2019} whilst study into their own potential topological characteristics has begun in ernest\cite{Jin:2017,Pan:2017,Song:2018}.

\begin{figure}
\centering
\begin{overpic}[width=\linewidth]{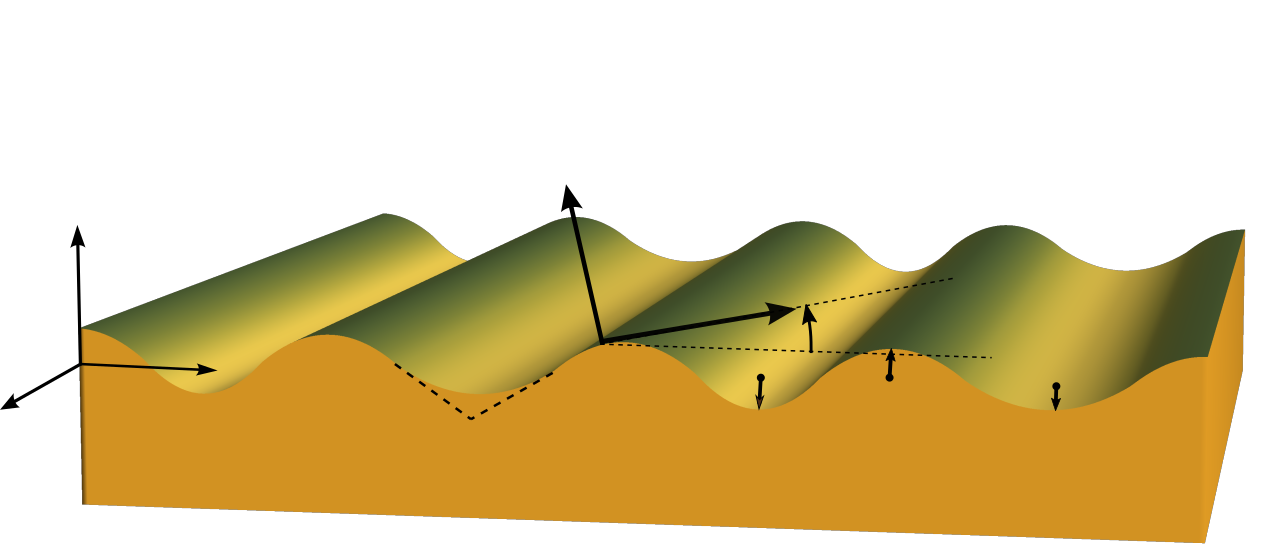}\put(89,3){(a)}
\put(15.5,15.5){$\bm{e}_x$}
\put(2,26){$\bm{e}_y$}
\put(-2,14){$\bm{e}_z$}
\put(60,21){$\bm{e}_\sigma$}
\put(46,28){$\bm{e}_\eta$}
\put(56,9){$a+t$}
\put(69,11.5){$a$}
\put(79,9){$a-t$}
\put(36.5,12){$\theta$}
\put(65,18){$\varphi$}
\put(8,22){$\epsilon_1$}
\put(8,10){$\epsilon_2$}
\end{overpic}
\begin{tikzpicture}
\draw (0,-1)
	--(.5,-1)
	--(.5,1.25)
	--(1.5,1.25)
	--(1.5,-1)
	--(3,-1)
	--(3,.75)
	--(5,.75)
	--(5,-1)
	--(6.5,-1)
	--(6.5,1.25)
	--(7.5,1.25)
	--(7.5,-1)
	--(8,-1);
\draw node at (-.1,2) [anchor=west] {(b)};
\draw[dotted] (1,1.75)--(7,1.75)--(7,-1.5)--(1,-1.5)--(1,1.75);
\draw[dotted] (-.25,0)--(8.25,0);
\draw[<->] (1,1.6)--(7,1.6) node at (4,1.6) [anchor=north] {$d$};
\draw[<->] (3,.85)--(5,.85) node at (4,.85) [anchor=south] {$v$};
\draw[<->] (0.5,1.35)--(1.5,1.35) node at (.85,1.35) [anchor=south] {$w$};
\draw[<->] (1.5,-.75)--(3,-.75) node at (2.25,-.75) [anchor=south] {$\delta$};
\draw[<->] (5,-.75)--(6.5,-.75) node at (5.75,-.75) [anchor=south] {$\delta$};
\draw[<->] (3.1,0)--(3.1,-1) node at (3.1,-.5) [anchor=west] {$V_0$};
\draw[<->] (4.9,0)--(4.9,.75) node at (4.9,.375) [anchor=east] {$V_3$};
\draw[<->] (1.4,0)--(1.4,1.25) node at (1.4,.625) [anchor=east] {$V_1$};
\draw node at (3.75,0) [anchor=south] {$V=0$};
\draw node at (1,1.75) [anchor=south] {$\sigma_0$};
\draw node at (1.5,-1) [anchor=north] {$\sigma_1$};
\draw node at (3,-1) [anchor=north] {$\sigma_2$};
\draw node at (5,-1) [anchor=north] {$\sigma_3$};
\draw node at (6.5,-1) [anchor=north] {$\sigma_4$};
\draw node at (7,1.75) [anchor=south] {$\sigma_5$};
\end{tikzpicture}
\caption{(Colour online) Panel (a): A general schematic of the biharmonic surface with curvilinear coordinates indicated. The local radius of curvature is defined as $R=(\partial_\sigma\varphi)^{-1}$. Panel (b): A diagram of the emergent Kronig-Penney model in $\sigma$ where $\delta=a(\pi-\theta)$, $w=(a+t)(\pi-\theta)$, $v=(a-t)(\pi-\theta)$, and $t<0$ such that of $v>w$ and $V_3<V_1$ with $t<0$ here.}
\label{fig:eKPmodel}
\end{figure}
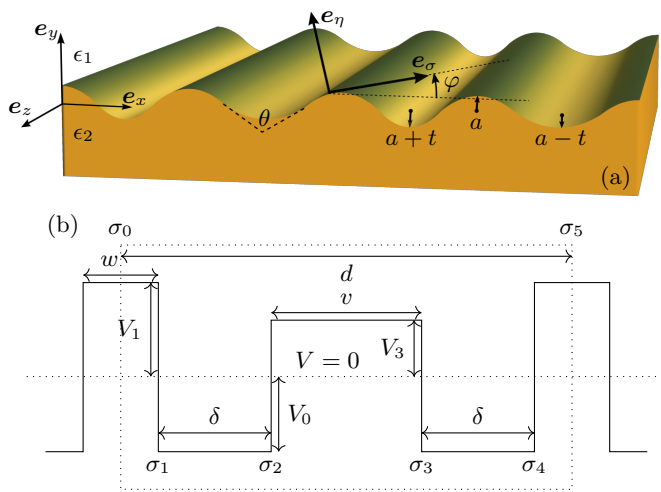

Herein, we consider an extended corrugated surface that is made to be bipartite in nature through a periodic variation of the radii of curvature between neighbouring peaks and troughs. The `bipartite-ness' here is constructed to effectively mimic the SSH model by keeping all peaks the same and varying the troughs between them. Such systems may be fabricated easily and with great accuracy using modern experimental techniques\cite{Kitson:1996,Lassaline:2020}.

\begin{figure}
\centering
\begin{overpic}[width=\linewidth]{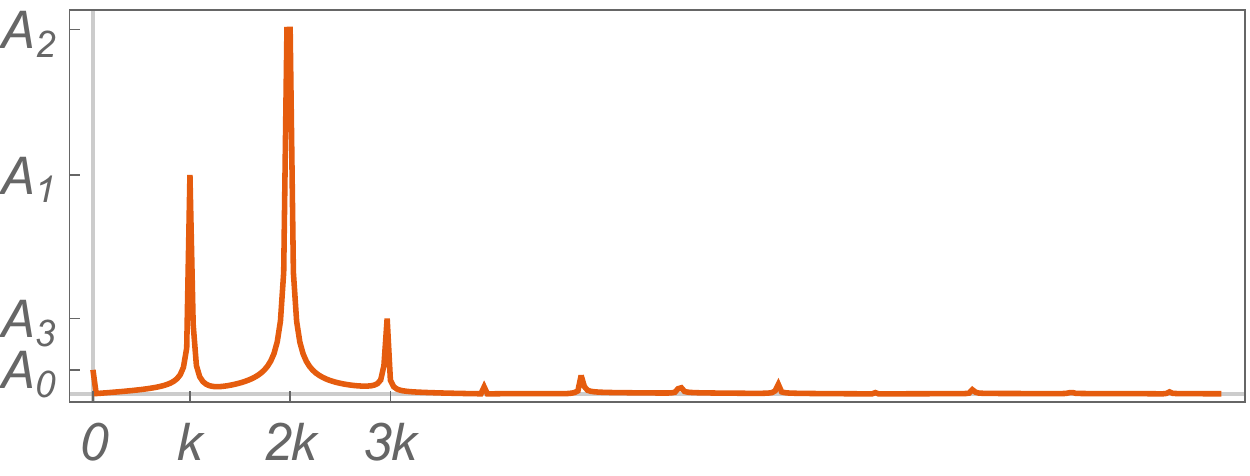}\put(94,32){(a)}
\end{overpic}
\begin{overpic}[width=\linewidth]{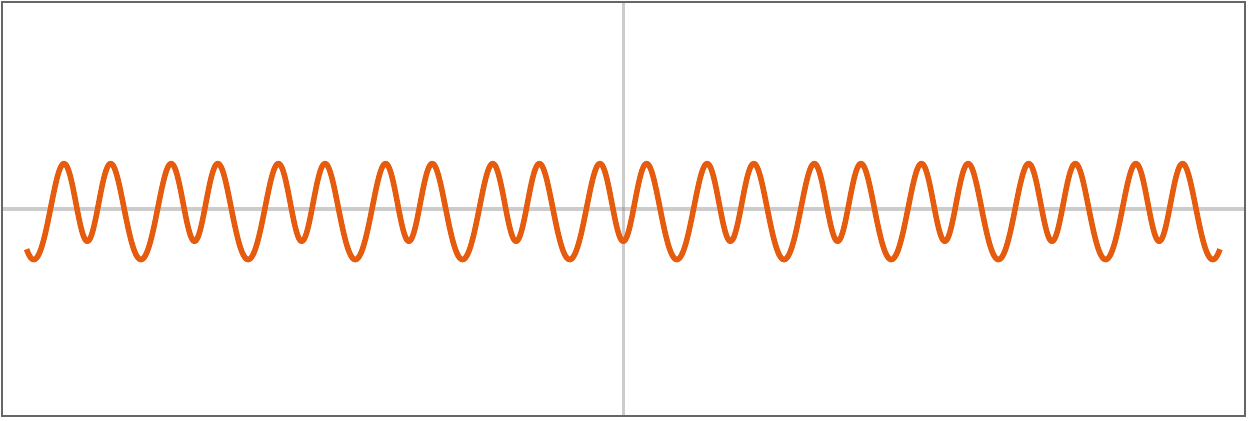}\put(94,28){(b)}
\end{overpic}
\caption{(Colour online) Panel (a): The Fourier transform of a typical grating profile shows three dominant harmonic modes in its construction. Panel (b): (Not to scale) A plot of the function $g(x)$ reveals that the bipartite corrugated structure is born from an underlying triharmonic pattern.}
\label{fig:gratingharmonics}
\end{figure}

By taking the Fourier transform of a typical bipartite grating, as shown in Fig.~\ref{fig:gratingharmonics}(a), it is clear that the dominant contributions to the periodic structure are the lowest three Fourier frequencies: $k$, $2k$, $3k$, with weights $A_{1,2,3}$ and the constant $A_0$, which is not crucial. Therefore, constructing the profile $g(x)=\sum_{n=0}^3A_n\cos(nkx)$ and plotting it as in Fig.~\ref{fig:gratingharmonics}(b) reveals that the bipartite structure is born from a triharmonic pattern. Note that panel (b) is not to scale, the widths of the corrugation are in fact much larger than the heights.


Within the first section, the emergent Kronig-Penney model is introduced. Its derivation is given in the supplementary material. The construction of the effective Schr\"odinger equation satisfied by the amplitude of the electromagnetic field follows Ref.~\cite{DellaValle:2010}. The bulk system is then solved in the following section. There, the bulk invariant is identified. Then the finite systems are solved and degenerate edge modes are observed in concordance with the bulk invariant. Finally, conclusions are drawn with respect to the results as presented.

\section{The Emergent Kronig-Penney Model}

As detailed within the supplementary material and in Ref.~\cite{DellaValle:2010}, the amplitude, $F(\sigma)$, of the SPP fields, $\{\bm{E},\bm{B}\}\propto F(\sigma)e^{[i(n_e+\Delta n)z-\gamma\eta]/\lambdabar}$, may be described through the following Schr{\"o}dinger-like equation:
\begin{equation}\label{eqn:KP}
-\Delta nF(\sigma)=-\frac{\lambdabar^2}{2n_e}\frac{\partial^2F}{\partial\sigma^2}+V(\sigma)F(\sigma),
\end{equation}
in the presence of a corrugated metal-dielectric interface, where $(\sigma,\eta)$ are curvilinear coordinates that are local to the surface as in Fig.~\ref{fig:eKPmodel}(a). In addition, $\lambdabar=\lambda(2\pi)^{-1}$ with $\lambda$ as the wavelength of the incident light ($\omega_i=c\lambdabar^{-1}$) that excites the SPP, $n_e$ is the effective index of the SPP and $\Delta n$ is a correction to this index induced by the perturbing `geometric' potential $V(\sigma)$ with:
\begin{equation}
n_e=\pm\sqrt{\frac{\epsilon_1\epsilon_2}{\epsilon_1+\epsilon_2}},\quad V(\sigma)=\frac{\lambdabar n_e}{2R(\sigma)}\sqrt{\frac{-1}{\epsilon_1+\epsilon_2}}.
\end{equation}
Here, $\epsilon_1$ is the dielectric constant of the dielectric above the metallic surface and $\epsilon_2=(n+i\kappa)^2$ is the dielectric function of the metal, where $n$ and $\kappa$ are the refractive index and extinction coefficients of the metal substrate at the given frequency $\omega_i$.


For the standard time-independent Schr{\"o}dinger equation (TISE) the eigenvalue is simply the energy of the state. Here, however, the energy of any SPP excitations is fixed at $\omega_i$ by the incident light. Instead, $\Delta n$ acts to alter the effective wavelengths of the SPPs. This is seen in the $z$-direction Fourier component of the fields: $e^{i(n_e+\Delta n)z/\lambdabar}$, where $e^{in_ez/\lambdabar}$ is the Fourier component in the absence of the grating. In the same way that a Schr{\"o}dinger particle has a Fourier time component of $e^{iEt/\hbar}$ and as such an energy-eigenvalue TISE, the current system has a Fourier `time' component of $e^{i\Delta nz/\lambdabar}$ and so an `energy' of $\Delta n$.

As such, the grating acts to create a non-zero $\Delta n$ and so modifies the effective wavelengths of the excited SPPs as $\lambda_{\rm SPP}=\lambda(n_e+\Delta n)^{-1}$. To be clear, a photon of wavelength $\lambda$ that interacts with the metal-dielectric grating will do so to generate an SPP excitation of effective wavelength $\lambda(n_e+\Delta n)^{-1}$.
Then, regardless of the sign of $n_e$, $\Delta n$ is a correction to this refractive index and so is not restricted to take only positive values, as is the case when considering Schr{\"o}dinger particles. $\Delta n$ is free to both increase or decrease the effective index of the resultant SPP mode. As a result, the band-spectra as presented herein, which are in terms of this correction $\Delta n$, are effectively dispersion spectra for the wavelength $\lambda_{\rm SPP}$ of the excited modes.

In \citet{DellaValle:2010}, the potentials of the peaks and troughs have the same magnitude since the magnitude of the local radius of curvature, $R(\sigma)=(\partial_\sigma\varphi)^{-1}$, never changes; only its sign.
We instead consider a system in which the local radius of curvature varies between peaks and troughs in not only sign but also magnitude. To mimic the SSH model, we take the radius of curvature of the peaks to be $R_\delta=-a$ whilst those of the troughs are taken to be $R_v=a-t$ and $R_w=a+t$ alternately and periodically where $t$ is a parameter that is free to vary within the range $1-a\leq t\leq a-1$. In this way, by ultimately modifying a single parameter $t$, the length of the unit-cell is maintained to be constant.

\section{The Primitive Unit-cell Solution}

The solution within the bulk then proceeds through the standard scattering formalism that is used to solve the Kronig-Penney model\cite{Kronig:1931}. The details of this calculation for the present system may be found in the supplementary material. The important result is that the simultaneous equations generated may be reduced to the following 2x2 unit-eigenvalue matrix equation:
\begin{equation}\label{eqn:Smat}
\begin{pmatrix}
r(k)&t(k)
\\
-t^*(k)e^{i\phi_k}&r^*(k)e^{i\phi_k}
\end{pmatrix}
\begin{pmatrix}
D_1
\\
C_3
\end{pmatrix}=
\begin{pmatrix}
D_1
\\
C_3
\end{pmatrix},
\end{equation}
where the matrix is denoted $S(k)$ and the matrix elements are given in the supplementary material. The transcendental equation that defines the energy bands may be found through $\det[S(k)-\mathbb{1}_2]=0$, which is solved numerically.

Since the present system has been reduced to an emergent Kronig-Penney model that is governed by the Schr{\"o}dinger equation, the topological information of the system resides in the Zak (one-dimensional Berry) phase of the wavefunction\cite{Raghu:2008,Hassani:2017,Marciani:2020,Zak:1989,Asboth:2016}, which in the present system is in fact the amplitude $F(\sigma)$:
\begin{equation}
\theta_Z=i\sum_m\int_{-\pi/d}^{+\pi/d}dk\bra{u_k^m}\ket{\partial_k^{}u_k^m},
\end{equation}
where $u_k^m(\sigma)=e^{-ik\sigma}F_k(\sigma)$ is the unit-cell periodic amplitude, the summation is over all `occupied' bands and the inner product signifies to integrate over the $\sigma$ coordinate within the unit-cell. Since bosons are considered here, the meaning of `occupied' is to say: all bands below the band gap.


In the finite system to be considered, two separate cases involving different physical parameters will be introduced. This is to best represent the topological edge/defect states. In both cases: $\lambda=\SI{.8}{\micro\metre}$, $a=\SI{8}{\micro\metre}$, and the aperture angle is $\theta=\SI{157}{\degree}$.

In one of the cases, the metal-dielectric is \ch{Ag}-\ch{SiO2} and so we take $\epsilon_1=3.9$ whilst the metal substrate is assumed to be pristine, which is valid since it is covered by the silicon-based dielectric thereby eliminating oxidisation effects. Using data from \citet{Jiang:2016} we then have that: $\epsilon_2=-27.6+0.919i$. 

In the other case of \ch{Au}-air, we take $\epsilon_1=1$ and the metal substrate is taken to be pristine also. Since gold does not readily oxidise, unlike silver, there is no issue with the dielectric above the substrate being air. Using data from \citet{Olmon:2012}: $\epsilon_2=-23.6+1.20i$.

As discussed in Ref.~\cite{DellaValle:2010}, due to the fact that $|{\rm Re}(\epsilon_2)|\gg|{\rm Im}(\epsilon_2)|$ in both of these above cases, the imaginary part of $\epsilon_2$ does not enter crucially into the asymptotic analysis and only appears, at leading order, as a term of the form $e^{-\xi|z|}$ with $\xi\in\mathbb{R}$. As such, it does not affect the profile, $F(\sigma)$, nor the behaviour of the SPP in the $\sigma$ coordinate.


Since the asymptotic analysis used to derive Eq.~(\ref{eqn:KP}) is based entirely upon the condition that $|R|\gg\lambdabar$\cite{DellaValle:2010}, the present extension respects this. However, the aperture angles $\theta$ of both the convex peaks and concave troughs must be made to be identical to ensure that the angle $\varphi(\sigma)$ does not vary discontinuously.

\begin{figure}
\centering
\begin{minipage}{.5\linewidth}
\begin{overpic}[width=\linewidth]{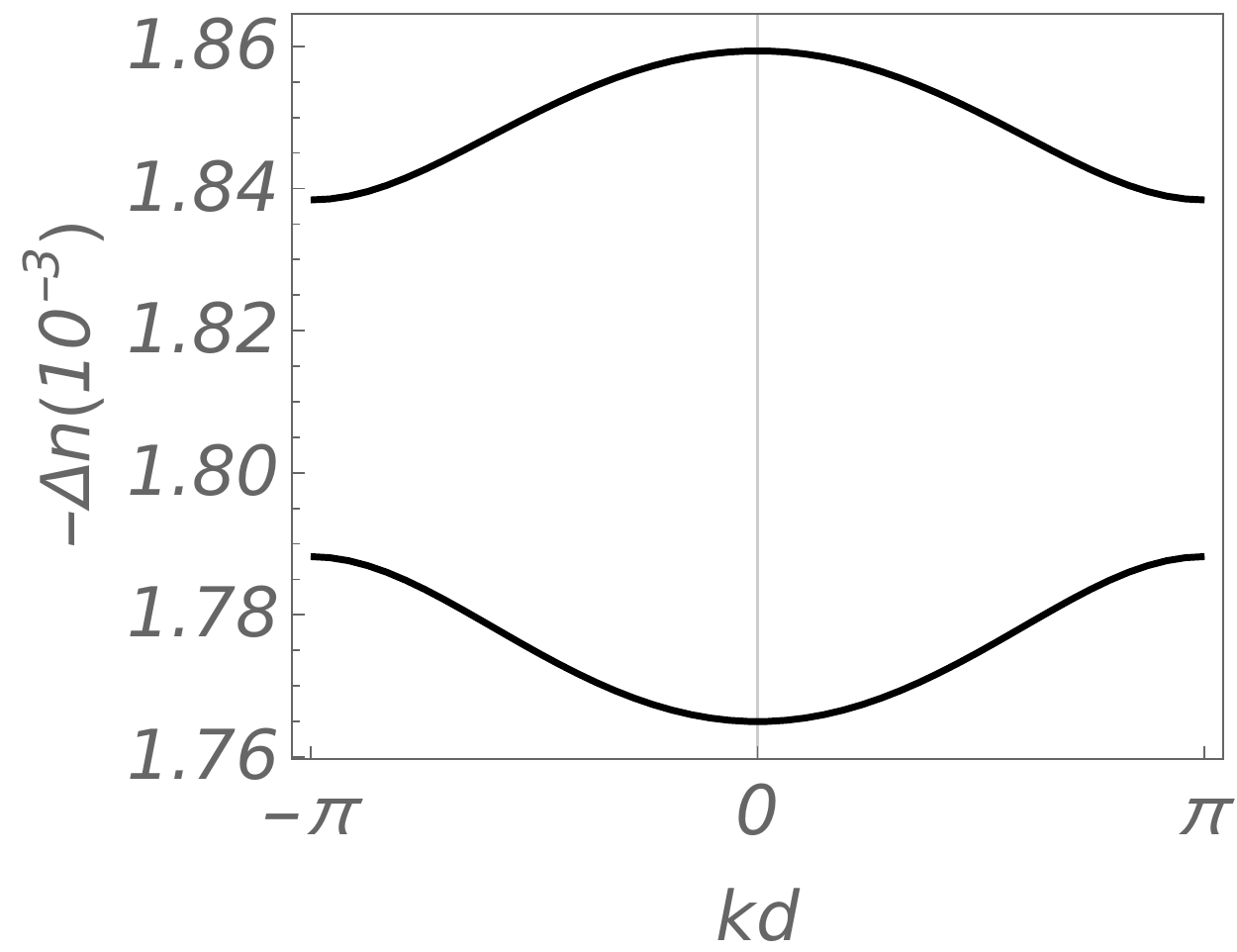}\put(25,68){(a)}
\end{overpic}
\end{minipage}%
\begin{minipage}{.5\linewidth}
\begin{overpic}[width=\linewidth]{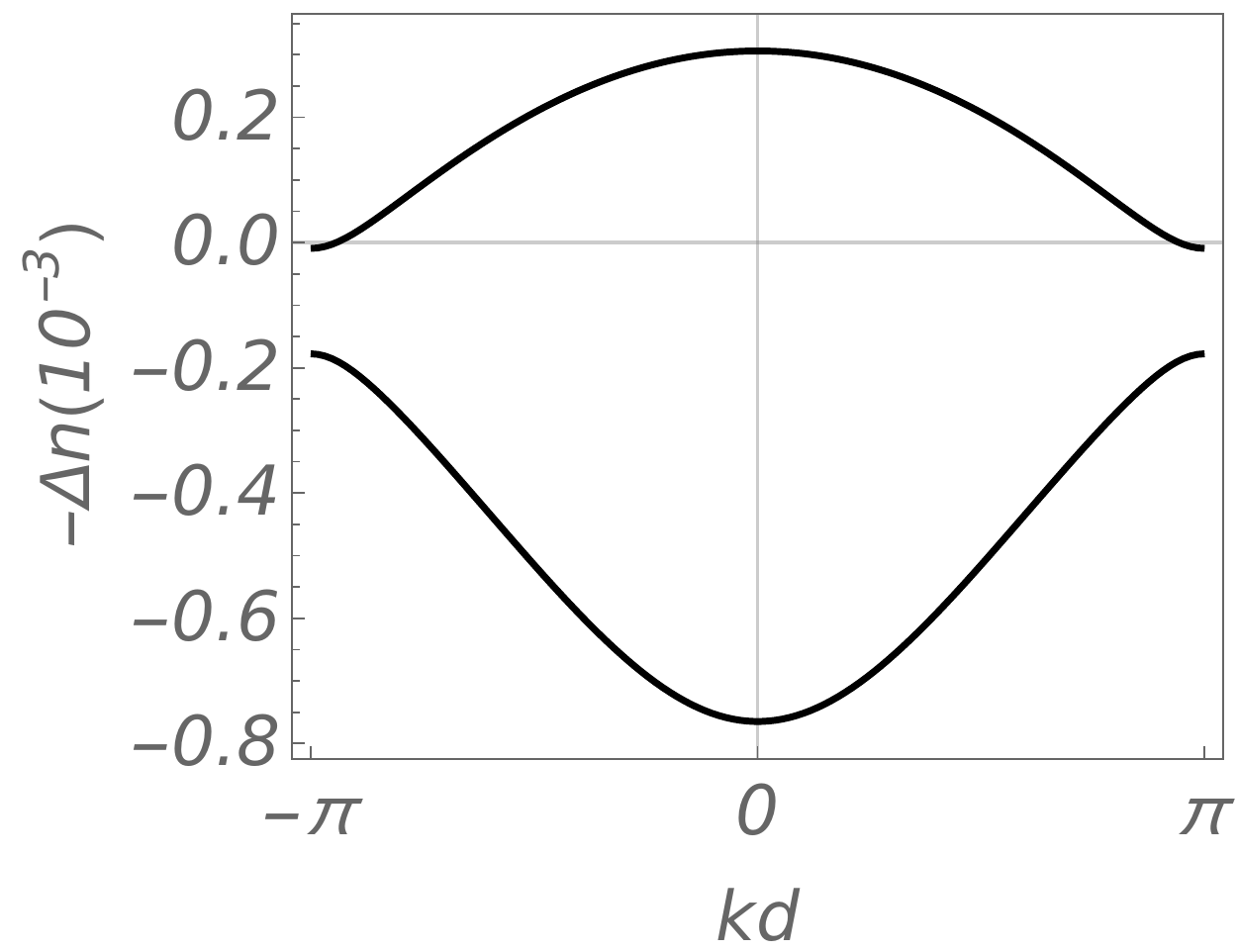}\put(25,68){(b)}
\end{overpic}
\end{minipage}
\begin{overpic}[width=\linewidth]{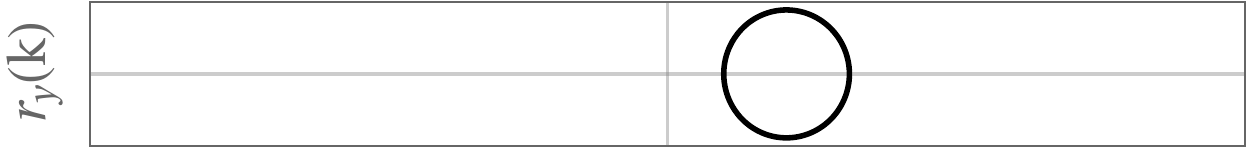}\put(8,8){(c)}\put(30,2){$\theta_Z=\pi$}
\end{overpic}
\begin{overpic}[width=\linewidth]{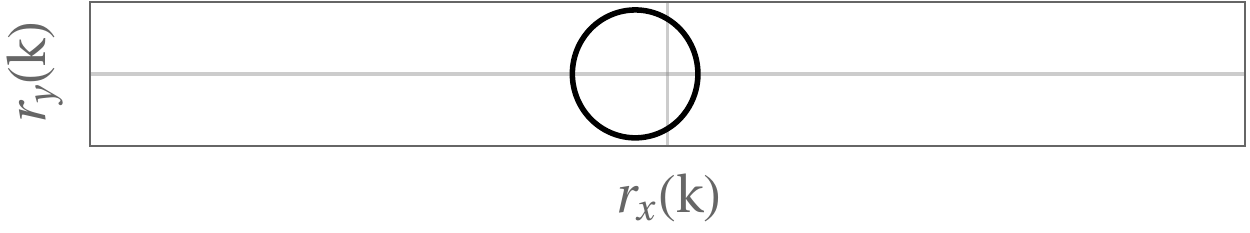}\put(8,14.5){(d)}\put(30,8){$\theta_Z=0$}
\end{overpic}
\caption{Panels (a,b): The band spectra for the bulk modes of the emergent Kronig-Penney model for the two cases with $t=\SI{2}{\micro\metre}$. Panel (a) \ch{Ag}-\ch{SiO2}, panel (b), \ch{Au}-air. Panels (b,c): The parametric windings of $r(k)=r_x(k)+ir_y(k)$ and the Zak phases for the lowest band of either case before and after the transition. Panel (c): $t=-\SI{1}{\micro\metre}$, panel (d): $t=\SI{1}{\micro\metre}$.}
\label{fig:bulk}
\end{figure}

Plots of the bulk bands may be seen in Figs.~\ref{fig:bulk}(a,b). Their similarity with those of the SSH model can be clearly seen. Indeed, the band gap closes at the transition point of $t=0$. Taking the lower bulk band of either case in Figs.~\ref{fig:bulk}(a,b), we may calculate the numerical Zak phase and find that $\theta_Z=\pi$ for $t<0$ and $\theta_Z=0$ for $t>0$. This step-wise change in the Zak phase invariant is also observed in the winding of the coefficient $r(k)$. Those of $r(k)$ for the lower band before and after the transition at $t=0$ are shown in Figs.~\ref{fig:bulk}(c,d). Thus the invariant that conforms with $\theta_Z$ truly is then $W=|1-W_r|$ where $W_r$ is the winding number of $r(k)$\cite{Smith:2019}.

We stress that the bands shown in Figs.~\ref{fig:bulk}(a,b) have a different nature with respect to those normally encountered in non-bipartite metal gratings. In those systems, usually a pair of low-lying bands are observed which bare resemblance to bonding and anti-bonding orbitals. This is clearly seen by analyzing the eigenstates corresponding to energies around the gap. In non-bipartite gratings, these are centred at different points of the unit cell, namely the (only) peak and the corresponding trough\cite{Barnes:1996}. This allows one to identify such bands as borne out of the linear superposition of both bound and excited states of the square potential wells generated by the grating. 

On the contrary, in the bipartite case the eigenstates around the gap have equal weight on the two peaks in the unit cell. The bipartite bands of Figs.~\ref{fig:bulk}(a,b) are therefore the result of the folding of the lowest band of a non-bipartite grating and are thus linear superpositions of only the bound states of square potential wells. A gap is opened by the unequal distances between peaks inside the cell and with peaks of neighbouring cells. When these become equal ({\it i.e.} at $t=0$), the grating is in practice not bipartite anymore, and therefore the gap {\it must} vanish, as discussed above.

\section{The Finite and Extended Unit-cell Solutions}

However, to find exponentially localised modes and/or observe any bulk-boundary correspondence, the finite system must be solved. The standard procedure is to terminate the edges of the lattice with hard walls such that the states are reflected back into the lattice at the edges. Such a hard termination may only be achieved perfectly with infinite-strength Dirac-delta potentials.

Physically, this may be approximated if the chain terminates at both ends with a trough that has both a large radius of curvature and a different dielectric deposited atop it, of constant $\epsilon_3$ where $\epsilon_3+\epsilon_2\sim0$. This has the effect of making this final barrier (in the K-P model) have both a large width, as a result of a large $R$, and a tall height, as a result of $\epsilon_3+\epsilon_2\sim0$. Silicon could constitute such a material since its dielectric constant has a value of $\epsilon_3\sim12$. The physical effect at play here is that the SPP, which is excited at $\omega_i<\omega_{\rm p}(1+\epsilon_1)^{-1/2}$, encounters the edge dielectric whereat $\omega_i\lesssim\omega_{\rm p}(1+\epsilon_3)^{-1/2}$ and so has its wavevector `shifted' along its dispersion. As a result, its group velocity tends to zero and thus it does not propagate meaningfully into the extended region.

Since, in the K-P model, the tunnelling probability of such a final barrier is many times smaller than those of the barriers of the bulk chain, it may be safely assumed that such a barrier may be approximated to have infinite strength. In a physical sense, the shear size of this final barrier acts to forbid the existence of SPPs within them and so it confines the SPPs within the region of interest. Theoretically, this may be realised effectively using hard walls. Then, due to the proximity of the final peaks to this effective infinitely positive potential, no states are permitted to reside therein. As a result, a system with $M$ peaks will permit $M-2$ states.

On the other hand, the opposite of the hard-wall would be an open boundary. In this case, the potential beyond the final well of the Kronig-Penney model ought to be zero, which is achieved by a flat surface in the physical system. Since a flat surface will always sustain SPPs\cite{Bliokh:2019}, such a boundary would cause the SPPs to leak out of the chain. Moreover, the abruptness of the change from the corrugation to the flat surface would no doubt introduce significant boundary effects at the resultant discontinuity. As such, open boundary conditions, like those used in typical tight-binding models, are simply not applicable here. Finally, the finite barrier with $\varepsilon_3$ could be used explicitly rather than through an infinite one. However this introduces the Shockley state\cite{Shockley:1939}, a separate entity, needlessly complicating the band spectra.



\begin{figure}
\centering
\begin{minipage}{.5\linewidth}
\begin{overpic}[width=\linewidth]{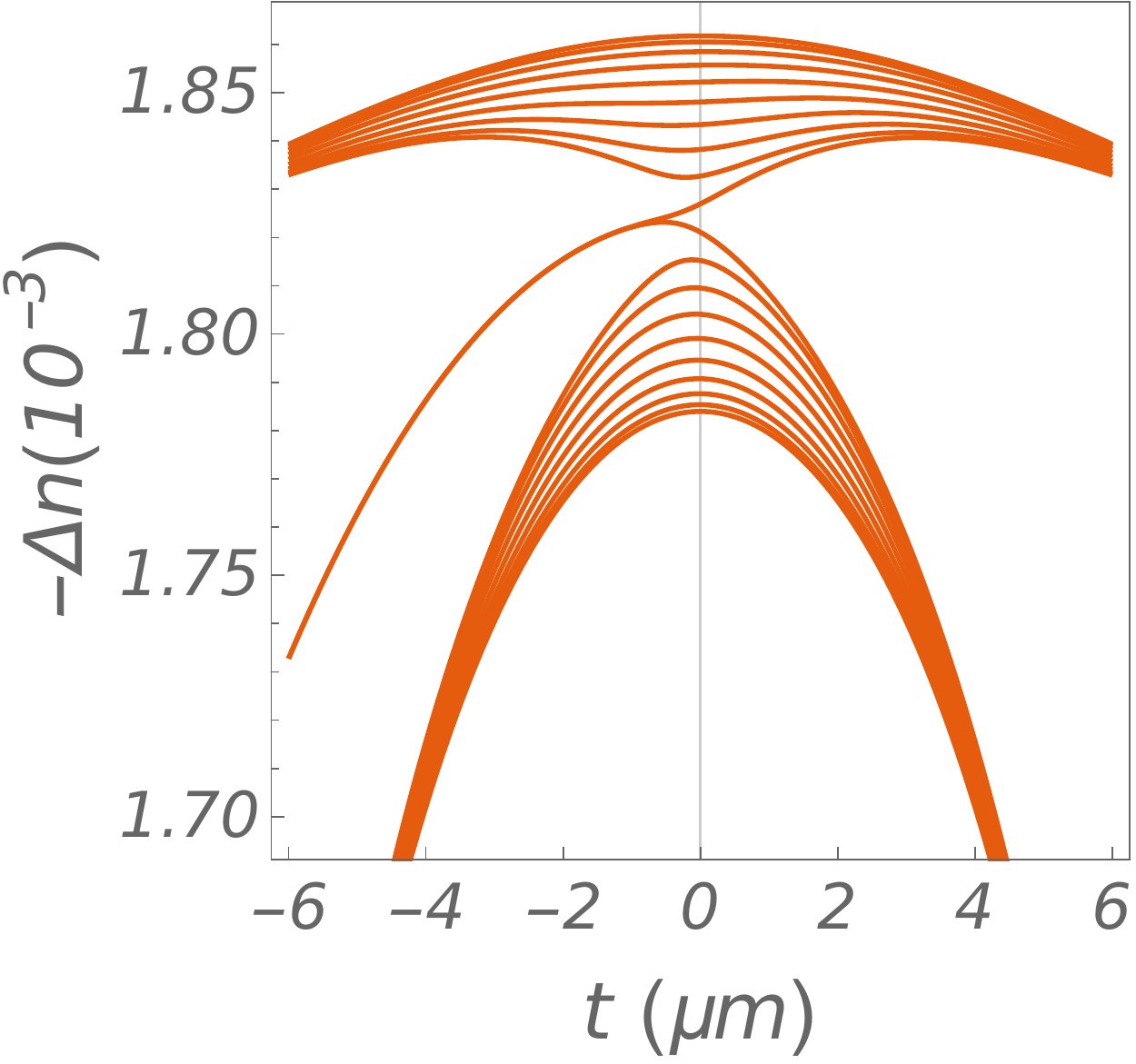}\put(25,60){(a)}
\end{overpic}
\end{minipage}%
\begin{minipage}{.5\linewidth}
\begin{overpic}[width=\linewidth]{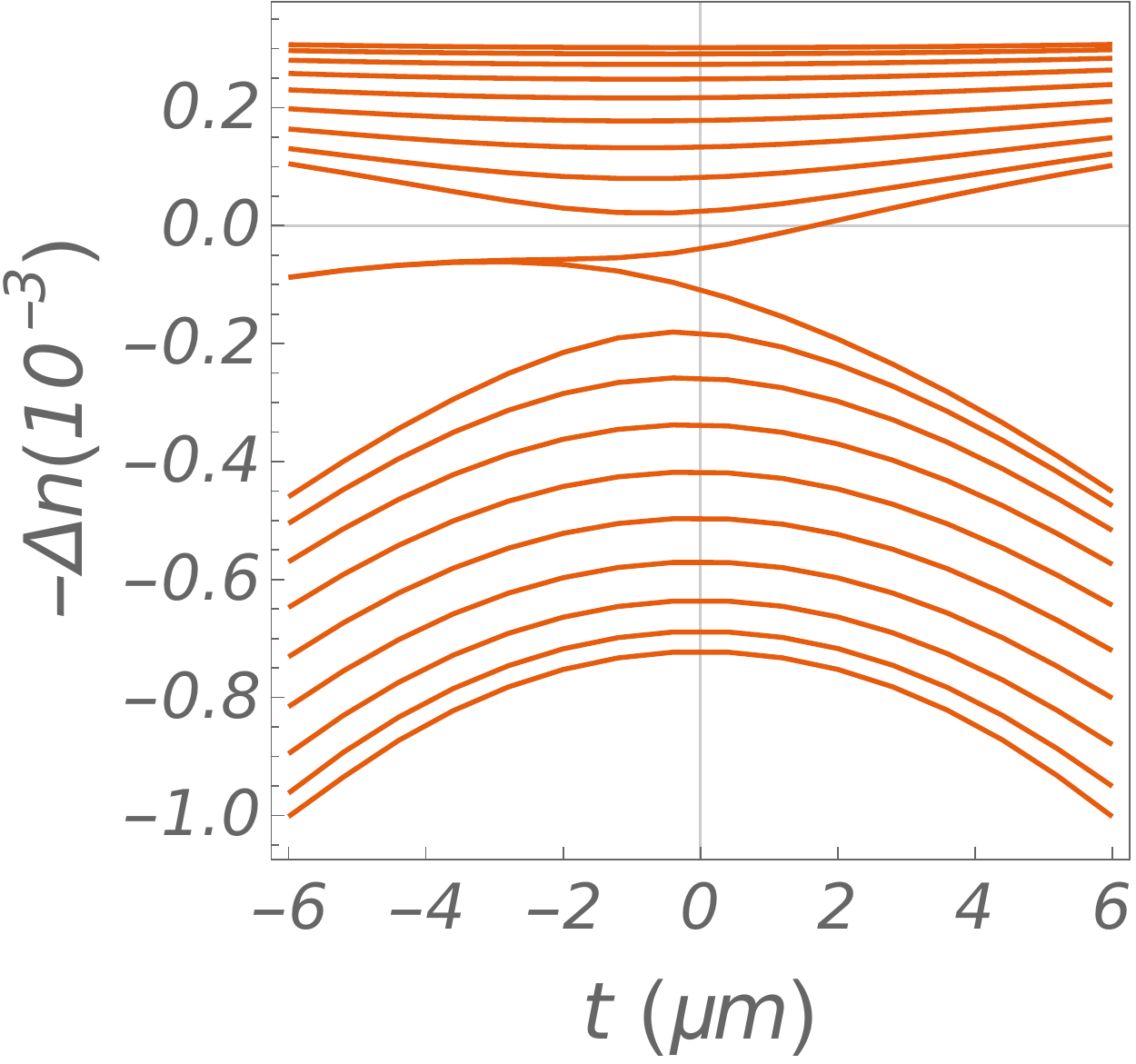}\put(25,60){(b)}
\end{overpic}
\end{minipage}
\begin{overpic}[width=\linewidth]{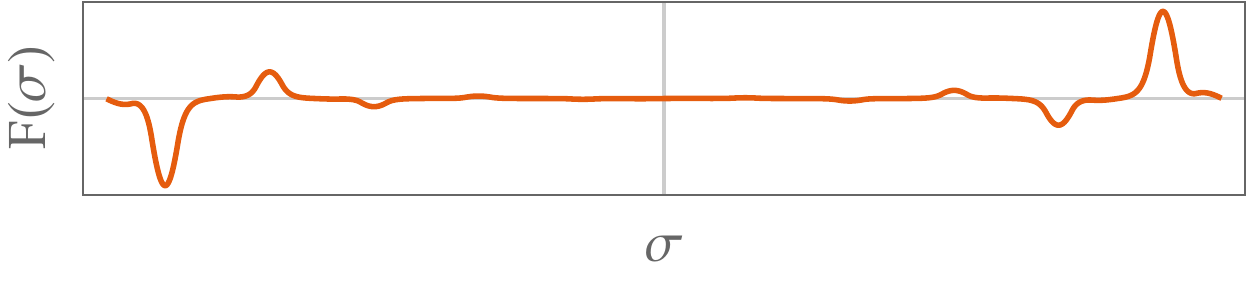}\put(7,19){(c)}
\end{overpic}
\begin{overpic}[width=\linewidth]{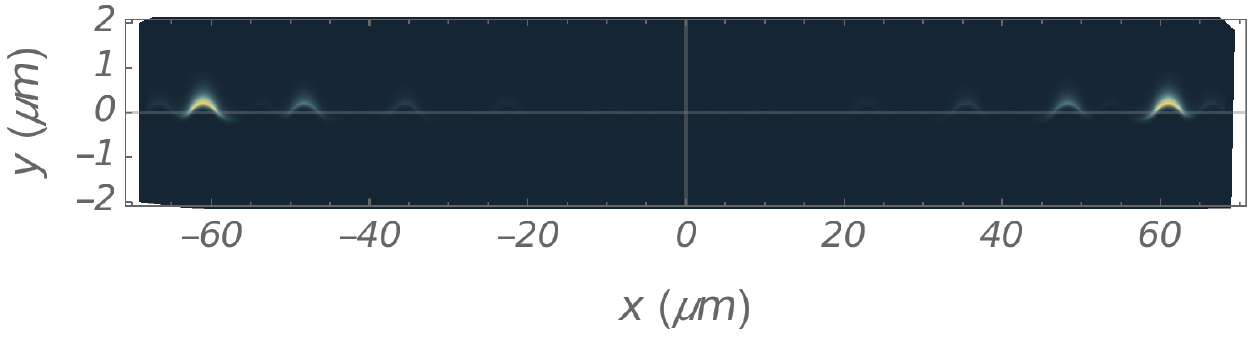}\put(11.5,22){\color{white}(d)}
\end{overpic}
\caption{(Colour on-line) Panels (a,b): The band spectra of $\Delta n$ vs $t$ for a chain of $44$ combined peaks and troughs terminating with hard walls. Panel (a): \ch{Ag}-\ch{SiO2}, panel (b): \ch{Au}-air. Although not clear, there are 20 bands in the plots. Panel (c): the 10th state for either system at $t=-\SI{3}{\micro\metre}$. Panel (d): the real-space magnetic field amplitude of this very mode.}
\label{fig:HW}
\end{figure}

In Figs.~\ref{fig:HW}(a,b) are shown the band spectra for this very case involving hard walls. As may be seen, mid-gap and degenerate edge modes may be seen for $t<0$ in conformity to the value of $\theta_Z$ within the bulk when $t<0$. Moreover, as may be seen in the behaviour of $F(\sigma)$ for these degenerate edge states in Figs.~\ref{fig:HW}(c,d), they are confined to a single sublattice only. As such, the edge states are topological for both the reasons of degeneracy and single-sublattice-confinement\cite{Asboth:2016}. 


As such, since the states: \begin{enumerate*}[label=(\roman*)] \item are degenerate, \item are confined to a single sublattice only, and \item conform with the bulk invariant $\theta_Z$\end{enumerate*}, then a bulk-boundary correspondence is established thereby protecting the states against chiral-symmetry-preserving lattice perturbations. This is true without an {\it a priori} knowledge of the presence of chiral symmetry within the system. Upon the calculation, its presence may be inferred from the character of the edge states. It is possible to construct a tight-binding model for such a system in order to confirm this and is done so elsewhere\cite{Smith:2020}.

A further way to observe exponentially localised modes is to consider a defect within the chain that is sufficiently far away from any other defect and/or the edges. 
As such, we consider an extended unit cell that contains a single
defect formed through the situation of two troughs of the same $R$ next to one another. Then, by taking a large enough number of peaks along with Bloch's theorem for the entire unit-cell, a localised mode may be seen to exist with an exponentially decaying profile at the defect.

\begin{figure}
\centering
\begin{minipage}{.5\linewidth}
\begin{overpic}[width=\linewidth]{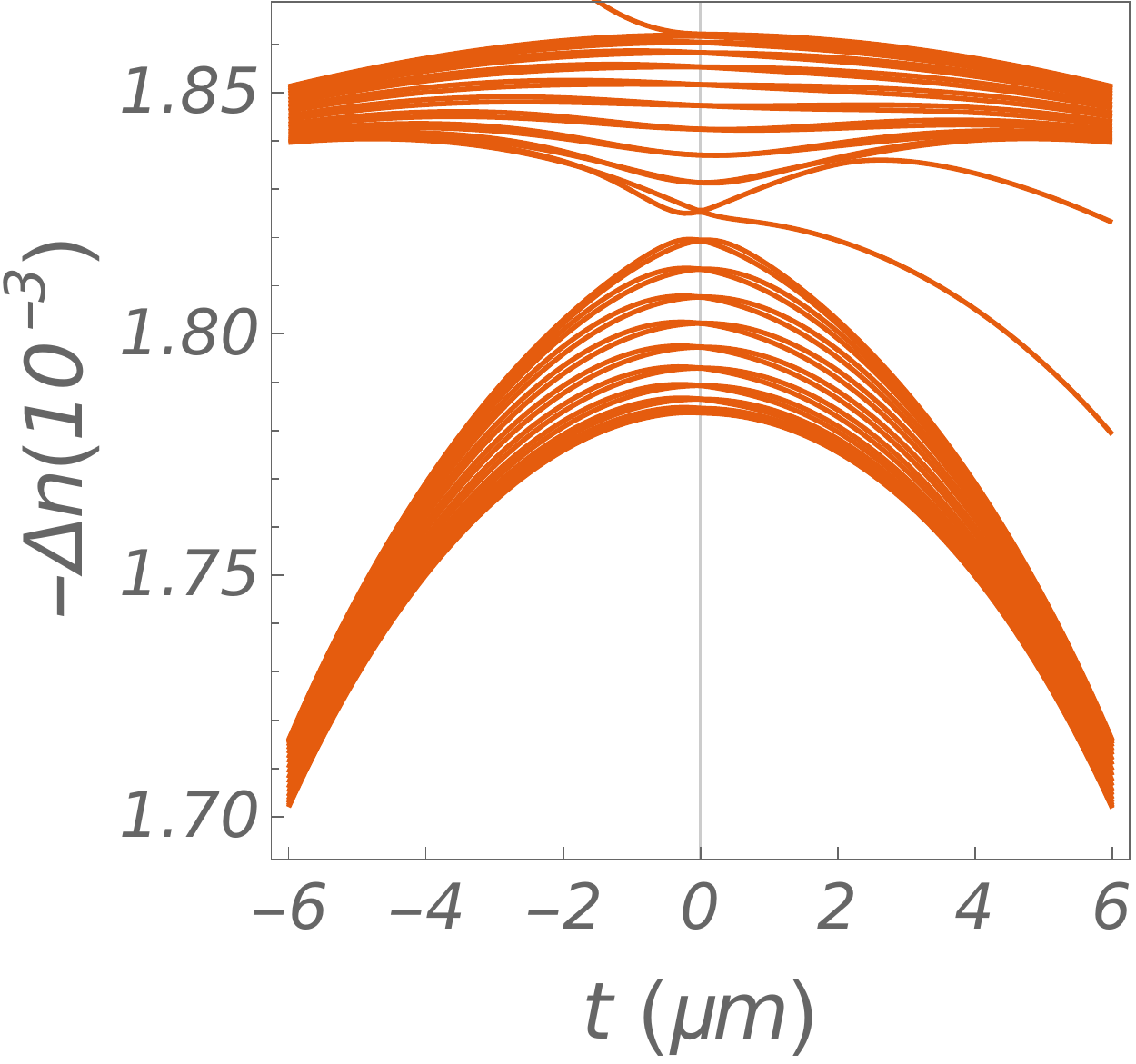}\put(25,74){(a)}
\end{overpic}
\begin{overpic}[width=\linewidth]{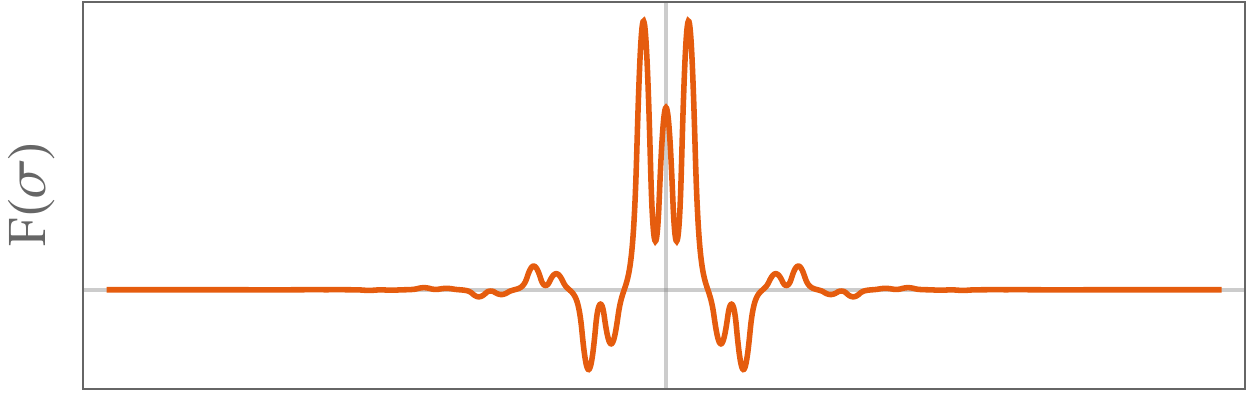}\put(7,24){(c)}
\end{overpic}
\begin{overpic}[width=\linewidth]{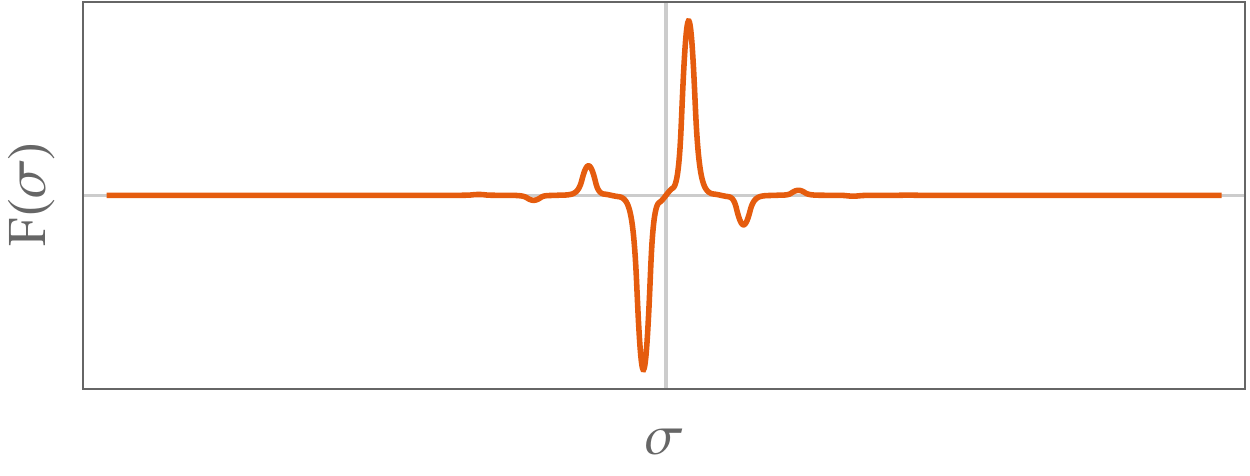}\put(7,30){(e)}
\end{overpic}
\end{minipage}%
\begin{minipage}{.5\linewidth}
\begin{overpic}[width=\linewidth]{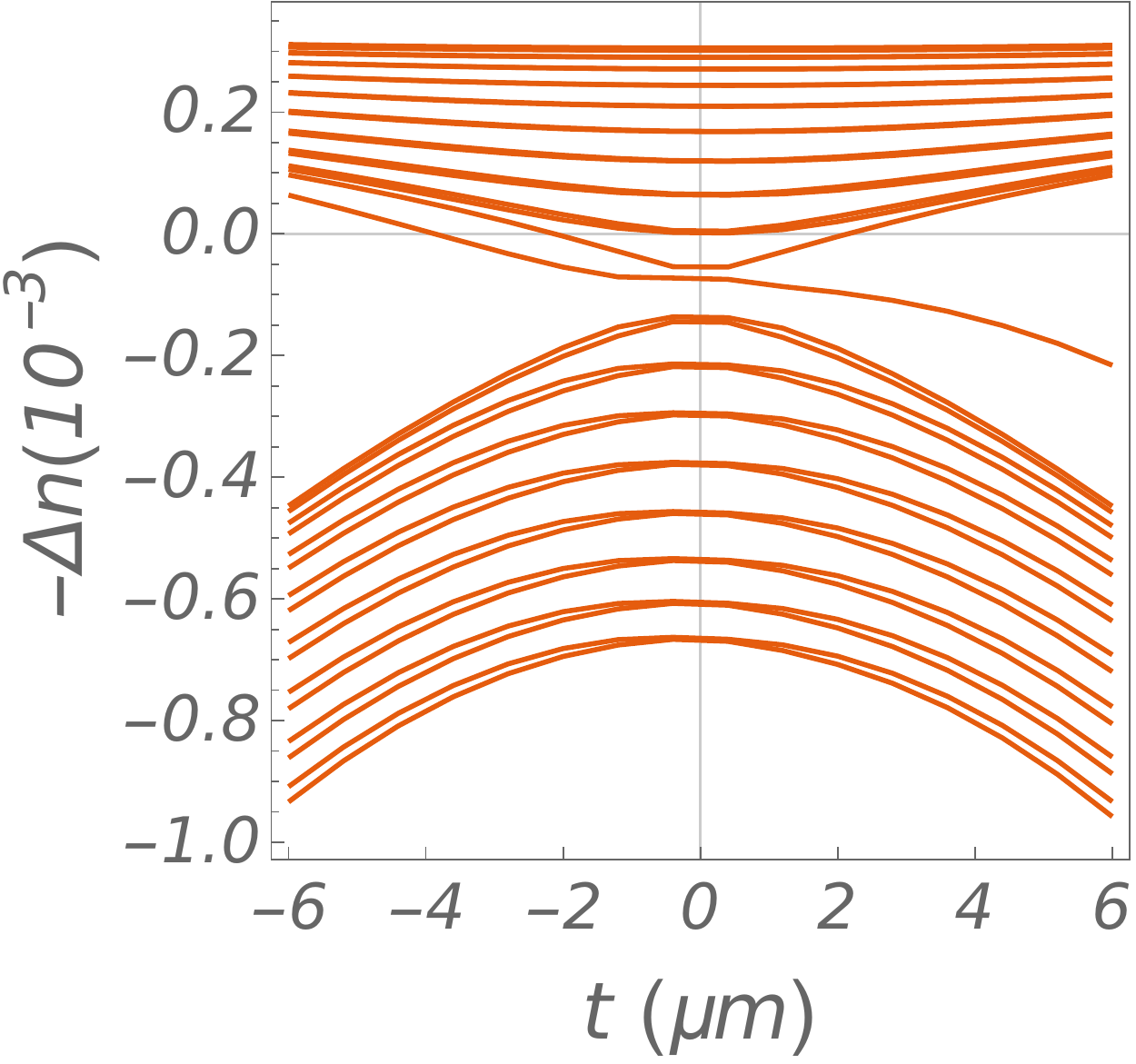}\put(25,67){(b)}
\end{overpic}
\begin{overpic}[width=\linewidth]{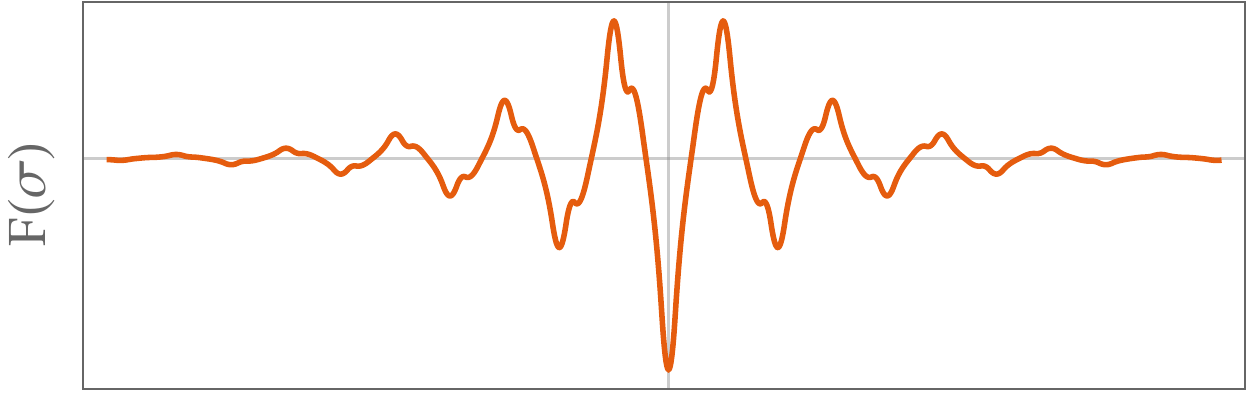}\put(8,24){(d)}
\end{overpic}
\begin{overpic}[width=\linewidth]{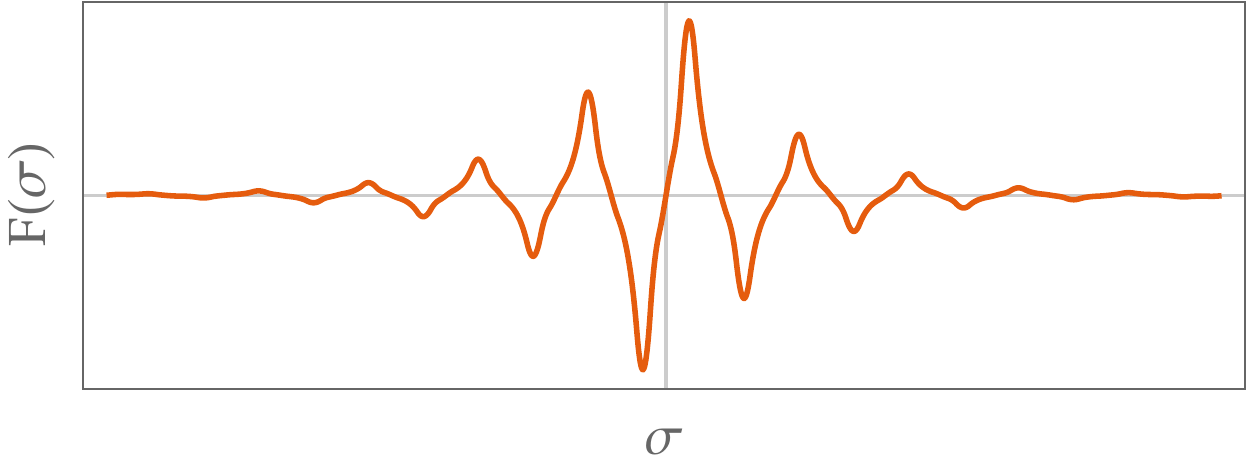}\put(8,30){(f)}
\end{overpic}
\end{minipage}
\caption{(Colour on-line) Panels (a,b): The band spectra of $\Delta n$ vs $t$ for a chain of $82$ combined peaks and troughs within the extended defect-including unit-cell. Panel (a): \ch{Ag}-\ch{SiO2}, panel (b): \ch{Au}-air. Panels (c,e): the 21st and 22nd states, respectively, of the \ch{Ag}-\ch{SiO2} system at $t=\SI{3}{\micro\metre}$. Panels (d,f): the 21st state of the \ch{Au}-air system for $t=\SI{-3}{\micro\metre}$ and $t=\SI{3}{\micro\metre}$, respectively. 
}
\label{fig:def}
\end{figure}

The band spectra of $\Delta n$ against $t$ for such an extended unit cell may be seen in Figs.~\ref{fig:def}(a,b).  As may be observed, there exist mid-gap, and therefore exponentially localised, defect states. Within the simple SSH model, there is always a single mid-gap topological defect mode within the entire phase space guaranteed by the step-wise change in $\theta_Z$ from $\pi$ to $0$ between the two chains\cite{Asboth:2016}.

However, here an interesting and highly non-trivial feature is that, as $t$ becomes more negative, the defect band tends towards the bulk bands. On the other hand, as $t$ becomes more positive, another mode appears within the bulk gap in the \ch{Ag}-\ch{SiO2} system. This extra state is indeed defect-localised however it is non-topological since it does not appear at the topological transition at $t=0$. Moreover, in this case, the defect state within the region $t<0$ approaches and joins the bulk much more rapidly compared to the \ch{Au}-air case. Finally, there is a defect state in the \ch{Ag}-\ch{SiO2} case above the upper band for $t<0$. However, since it is not within the gap, it cannot be a topological state and this is shown (if plotted) by its weight upon both sublattices.

All these observations are to say that the defect is capable of being constructed in such a way that its localisation character is affected in a topologically trivial way. The local potential can become large enough, relative to the bulk, to forbid the state even though the system is topologically non-trivial.

In both systems, 
when $t<0$, the potentials that neighbour the defect, $V_3\propto(a-t)^{-1}$, are reduced from their value at $t=0$ whereas when $t>0$ the potentials are greater. As such, the localisation of the defect is reduced when $t<0$ and enhanced when $t>0$. Then, the state within $t<0$ `leaks out' of the defect to become a bulk mode whilst another state is allowed to exist within $t>0$, which are respectively due to the decreased and increased apparent depths of the defect well in either case. This is an unfortunate state of affairs that may not be avoided with such a defect.

This indicates, rather nicely, a condition in which a symmetry-protected state is forbidden. The difference between the \ch{Ag}-\ch{SiO2} and \ch{Au}-air systems is in the depths and heights of the wells and barriers of their respective Kronig-Penney models arising from their values for $V_{0,1,3}\propto[-(\epsilon_1+\epsilon_2)]^{1/2}$. In the case of \ch{Au}-air, the changing potential environment of the defect is less pronounced than in the \ch{Ag}-\ch{SiO2} case due to the smaller potential magnitudes in the former case compared to the latter case. Within a prototypical tight-binding model of such a system\cite{Smith:2020} these modified potential environments act to generate an on-site potential at the defect that is greater in magnitude to that of the bulk. Therefore the states are energetically forbidden from residing there and become bulk modes as a result.

If, instead, the defect is 
formed through a dielectric deposition technique akin to that which generates hard walls in the first considered case, then identical topological edge states that were observed there would exist here. They would be unlike standard defect states, however, as they would localise upon either side of the large defect trough in identical fashion to how the edge states in the first case localise to one side of the edge and not at all within the region defined by the extra deposited dielectric $\epsilon_3$. In analogy to the SSH model as applied to polyacetalene, this would be akin to forming a defect from a carbon-carbon triple bond.

However, and crucially, the {\it edge states}, which arise when hard wall boundary conditions are imposed, are degenerate and have weights on a single sublattice only {\it in both cases}. As such, they are still symmetry-protected. It is only the defect mode in the \ch{Ag}-\ch{SiO2} case that loses its protection within a region of the phase space due to a topologically trivial boundary effect wherein the potential at the defect is greater than that within the bulk. Yet, within the other region of the phase space it still exhibits single-sublattice-confinement.

\section{Discussion}

Within a tight-binding calculation, there are on-site potentials at each lattice site (peak) that transpire to be identical within the bulk\cite{Smith:2020}. Their strength not only depends on the depth of the potential well formed from the peak but also on the neighbouring potential barriers formed by the troughs. 

In the case of the hard wall, the peak that precedes the infinite potential is neighboured to the right(left) by a trough of $|R|=a+t$ and said infinite potential to the left(right). As such it has an on-site energy that is different, but crucially lesser, than in the bulk where each peak is neighboured by troughs with $|R|=a-t$ and $|R|=a+t$. When $t\to-\SI{6}{\micro\metre}$ then this on-site energy starts to affect the edge state. Yet it remains degenerate, as seen in the bands, and confined to a single sublattice {\it in the bulk of the chain} away from the edges. It is only its edge behaviour that is affected. This is a testament to the robustness of the state. However, in the case of the defect state, it is neighboured by $|R|=a+t$ on both sides and so has a different on-site potential than in the bulk. When $t<0$ this potential is greater and transpires to force the state away from the defect and into the bulk. When $t>0$ it is lesser so causes an increased localisation of the state to the defect and, moreover, the admittance of a further non-topological state when $t\to+\SI{6}{\micro\metre}$.

The present theory is based upon an approximation: that of $|R|\gg\lambdabar$. This holds for the parameters as used since we have that $|R|$ is at most $\SI{14}{\micro\metre}$ and at least $\SI{2}{\micro\metre}$, both of which are much larger than $\lambdabar=\SI{800}{\nano\metre}(2\pi)^{-1}\approx\SI{0.1}{\micro\metre}$. In Ref.~\cite{DellaValle:2010}, a full numerical calculation was conducted, which found that the approximation holds well for both $\SI{1}{\micro\metre}\leq a\leq\SI{20}{\micro\metre}$ and $\SI{350}{\nano\metre}\leq \lambda\leq\SI{1000}{\nano\metre}$.

\section{Conclusion}

Topological symmetry-protected localised SPP modes have been predicted to exist upon biharmonic {\bf (tri-harmonic?)} metal-dielectric gratings by solving an emergent Kronig-Penney model found from Maxwell's equations. The necessary bulk-boundary correspondence was established between the Zak phase bulk invariant $\theta_Z$ and the number of edge modes. The presence of chiral symmetry was then inferred from the solution thereby showing the states to be topologically protected.

However, interesting boundary/edge effects were observed that destroyed the protection of the defect states thereby indicating one such limit in which the bulk-boundary correspondence breaks down in one-dimension through a symmetry-breaking procedure, which has not been artificially introduced but instead arises naturally from the physical system under consideration.

Given the nature of the derivation of the Schr{\"o}dinger-like equation through the asymptotic expansion and its applicability to such corrugated surfaces as presented, an extension of the work could be to consider a Moir{\'e} superlattice\cite{Senlik:2009,Kocabas:2009}. In this situation, the heights of the peaks and troughs vary in size through some sinusoidal envelope where the present approximation that $|R|\gg\lambdabar$ would remain applicable. In such systems, localised states may be observed and, due to their similarity to the present system and the applicability of the current model therein, such states may also be topological in nature.

\section*{Acknowledgements}
T.B.S. acknowledges the support of the EPSRC through the Ph.D. studentship grant EP/N509565/1. The authors acknowledge support from the Royal Society International Exchange grant IES\textbackslash R3\textbackslash 170252 and the use of the freely available RootSearch.m Mathematica package developed by Ted Ersek\footnote{\url{https://library.wolfram.com/infocenter/Demos/4482}}.

\section*{Conflicts of interest}
There are no conflicts to declare.




\bibliography{bibliography}
\bibliographystyle{rsc} 

\onecolumngrid
\section{Supplementary Material}

\subsection{The Generic Scattering Solution}

The details of the bulk unit cell, as shown in Fig.~\ref{fig:eKPmodel}, may be described. Firstly, it is constructed to be centred upon the trough that has $R=a-t$, thereby making the unit cell centro-symmetric, and so:
\begin{equation}
V_0=-\frac{\lambdabar n_e}{2a}\sqrt{\frac{-1}{\epsilon_1+\epsilon_2}},\quad V_1=-\frac{\lambdabar n_e}{2(a+t)}\sqrt{\frac{-1}{\epsilon_1+\epsilon_2}},\quad V_3=-\frac{\lambdabar n_e}{2(a-t)}\sqrt{\frac{-1}{\epsilon_1+\epsilon_2}},
\end{equation}
Since the distances between the interfaces in  $\sigma$ are the arc-lengths of the curved surfaces between the edges of the peaks/troughs, the widths of the wells (peaks) in $\sigma$ are both $\delta=a(\pi-\theta)$ whilst the widths of the barriers (troughs) in $\sigma$ are $w=(a+t)(\pi-\theta)$ and $v=(a-t)(\pi-\theta)$. 

Now we exactly solve the presented system by requiring that the amplitude $F(\sigma)$ and its first derivative with respect to $\sigma$ be continuous at the potential interfaces. The standard solution of the TISE is that the amplitude (wavefunction) is a superposition of right and left moving waves within each well and barrier. So, explicitly: $F_k(\sigma)=\sum_{j=1}^N\theta(\sigma-\sigma_{j-1})\theta(\sigma_j-\sigma)f_{j,k}(\sigma)$, where $N$ is the number of regions within the unit cell, which is five in our bulk unit cell, $k$ is the Bloch wavevector that is defined through $F_k(\sigma+d)=F_k(\sigma)e^{ikd}$.

The amplitudes within each distinct region are given by: $f_{j,k}(\sigma)=C_je^{iq_j(k)\sigma}+D_je^{-iq_j(k)\sigma}$, where: $q_j(k)=\lambdabar^{-1}[2n_e(-\Delta n(k)-V_j)]^{1/2}$, and we have that $q_2=q_4\equiv q_0$ and $q_5\equiv q_1$. The dependence of the wavevectors $q_j(k)$ upon the Bloch wavevector $k$ is dropped from here on.

To solve the Kronig-Penney model in the standard way, the amplitude $F(\sigma)$ and its first derivative in $\sigma$ are assumed to be continuous at every interface. Thus at the $i$th interface we have:
\begin{equation}
f_{i,k}(\sigma_i)-f_{i+1,k}(\sigma_i)=0,\quad [\partial_\sigma f_{i,k}]](\sigma_i)-[\partial_\sigma f_{i+1,k}](\sigma_i)=0,
\end{equation}
which, when evaluated, leads to the transfer matrix equation:
\begin{equation}
\begin{pmatrix}
C_{i+1}\\D_{i+1}
\end{pmatrix}=
M_i\begin{pmatrix}
C_i\\D_i
\end{pmatrix},
\end{equation}
with the Bloch condition requiring that $\{C_5,D_5\}=\{C_1e^{-iq_1d},D_1e^{iq_1d}\}e^{ikd}$. There are four interfaces yielding eight boundary conditions and one Bloch condition yielding a further two. Thus, all ten coefficients are completely specified and through simple but lengthy manipulations, the following matrix equation may be arrived at:
\begin{equation}
S(k)
\begin{pmatrix}
D_1
\\
C_3
\end{pmatrix}=
\begin{pmatrix}
r(k)&t(k)
\\
-t^*(k)e^{i\phi_k}&r^*(k)e^{i\phi_k}
\end{pmatrix}
\begin{pmatrix}
D_1
\\
C_3
\end{pmatrix}=
\begin{pmatrix}
D_1
\\
C_3
\end{pmatrix},
\end{equation}
where:
\begin{align}
r(k)&=\left(\frac{-4q_0^2q_1q_3e^{ikd}+\left[(q_0^2-q_1q_3)\sin(q_0\delta)-iq_0(q_1-q_3)\cos(q_0\delta)\right]^2e^{i(q_1v-q_3w)}}{\left[(q_0^2+q_1q_3)\sin(q_0\delta)+iq_0(q_1+q_3)\cos(q_0\delta)\right]^2}\right)
e^{i(q_1v+q_3w)},
\\
t(k)&=\frac{-2iq_0q_3e^{i(q_3-q_1)v/2}e^{iq_1(v-\delta)}}{\left[(q_0^2+q_1q_3)\sin(q_0\delta)+iq_0(q_1+q_3)\cos(q_0\delta)\right]^2}\times\nonumber
\\
&\left(\left[(q_0^2-q_1q_3)\sin(q_0\delta)-iq_0(q_1-q_3)\cos(q_0\delta)\right]e^{-ikd}\nonumber
-\left[(q_0^2-q_1q_3)\sin(q_0\delta)+iq_0(q_1-q_3)\cos(q_0\delta)\right]e^{i(q_3w-q_1v)}\right),
\\
e^{i\phi_k}&=e^{i(q_1v+q_3w)}\left(\frac{\left[(q_0^2+q_1q_3)\sin(q_0\delta)-iq_0(q_1+q_3)\cos(q_0\delta)\right]^2}{\left[(q_0^2+q_1q_3)\sin(q_0\delta)+iq_0(q_1+q_3)\cos(q_0\delta)\right]^2}\right).\end{align}
The non-trivial solution of this matrix equation yields the transcendental equation through:
\begin{equation}
\det[S(k)-\mathbb{1}_2]=\det[S(k)]-\tr[S(k)]+1=[1-r^*(k)]e^{i\phi_k}+1-r(k)=0,
\end{equation}
since $|r|^2+|t|^2=1$, which may be shown, thereby making $S(k)$ unitary since $|{\rm det}[S(k)]|=1$. In full, and without pointless algebra, the transcendental equation is:
\begin{align}\label{eqn:treqn}
4q_0^2q_1q_3\cos(kd)&=\left[(q_0^2-q_1q_3)^2\sin^2(q_0\delta)-q_0^2(q_1-q_3)^2\cos^2(q_0\delta)\right]\cos(q_1w-q_3v)\nonumber
\\
&-\left[(q_0^2+q_1q_3)^2\sin^2(q_0\delta)-q_0^2(q_1+q_3)^2\cos^2(q_0\delta)\right]\cos(q_1w+q_3v)\nonumber
\\
&+\left[q_0(q_1-q_3)(q_0^2-q_1q_3)\sin(q_1w-q_3v)-q_0(q_1+q_3)(q_0^2+q_1q_3)\sin(q_1w+q_3v)\right]\sin(2q_0\delta).
\end{align}
Due to the intricate relation of $\Delta n$ to $k$ through the wavevectors $q_j$, this may only be solved numerically.

%
%
%

\subsection{Derivation of the Schr{\"o}dinger-like Equation}

The covariant formulation of Maxwell's equations in a general curved spacetime in the presence of dielectrics and the absence of free charges and currents is given by:
\begin{equation}\label{eqn:covEM}
\partial_\nu {\cal D}^{\mu\nu}=0,
\quad
F_{[\mu\nu;\lambda]}=0,
\end{equation}
where $F_{\mu\nu}$ is the contravariant electromagnetic field tensor and ${\cal D}^{\mu\nu}$ is the covariant electromagnetic displacement tensor. In general, the Bianchi identity is in terms of covariant derivatives ($;\gamma=\nabla_\gamma$), however due to the symmetry and anti-symmetry of the electromagnetic field tensor and the Christofel symbols in their lower indices respectively, the covariant derivatives reduce to standard Minkowski derivatives, and so:
\begin{equation}
F_{[\mu\nu;\lambda]}=F_{[\mu\nu,\lambda]}=\frac{1}{3}\left(\partial_\lambda F_{\mu\nu}+\partial_\nu F_{\lambda\mu}+\partial_\mu F_{\nu\lambda}\right)=0.
\end{equation}
On the other hand, the electromagnetic displacement tensor is defined in terms of the field tensor, the metric tensor, and the polarisation tensor as:
\begin{equation}
{\cal D}^{\mu\nu}=\frac{\sqrt{-g}}{c}g^{\mu\alpha}F_{\alpha\beta}g^{\beta\nu}-{\cal M}^{\mu\nu}=\frac{\sqrt{-g}}{c}g^{\mu\alpha}\tilde{F}_{\alpha\beta}g^{\beta\nu}.
\end{equation}
where the polarisation tensor has been absorbed into $F_{\alpha\beta}$ to form $\tilde{F}_{\alpha\beta}$ as a modified electromagnetic tensor with dielectric prefactors before the electric fields. As a result of this Lorentz-covariant formulation, Maxwell's equations may be recast in a completely covariant way in any coordinate system we choose so long as we know the metric tensor of said coordinate system.

In the case of the corrugated surface considered herein, we introduce the set of curvilinear coordinates that are defined locally to the surface of $(\sigma,\eta,z)$, where $\sigma$ is the direction along the surface, {\it i.e.} the arc-length, and $\eta$ is the coordinate perpendicular to the surface. At each point in $\sigma$ along the surface, the curvilinear coordinates are related to the Cartesian coordinates by:
\begin{equation}\label{eqn:xyTse}
\begin{pmatrix}
x
\\
y
\end{pmatrix}=
\begin{pmatrix}
\cos[\varphi(\sigma)]&-\sin[\varphi(\sigma)]
\\
\sin[\varphi(\sigma)]&~~\cos[\varphi(\sigma)]
\end{pmatrix}
\begin{pmatrix}
\sigma
\\
\eta
\end{pmatrix},
\end{equation}
where $\varphi(\sigma)$ is shown in Fig.~\ref{fig:eKPmodel}(a), is defined through $\bm{e}_x\cdot\bm{e}_\sigma=\cos[\varphi(\sigma)]$, and depends explicitly upon $\sigma$. This dependence will be dropped from now on for brevity.

The metric tensor, $g_{\mu\nu}$, in a general spacetime satisfies:
\begin{equation}
(ds)^2=g_{\mu\nu}dx^\mu dx^\nu,
\end{equation}
where $ds$ is the infinitesimal spacetime arc-length and $dx^\mu=(cdt,dx,dy,dz)$ is the infinitesimal 4-position, with a metric equivalent to the Minkowski metric (in the astrophysical convention) of $\eta_{\mu\nu}={\rm diag}(-1,1,1,1)$, thus defining the arc-length in Minkowski space of: $(ds)^2=-(cdt)^2+(dx)^2+(dy)^2+(dz)^2$. 

Using the relationship between $(x,y)$ and $(\sigma,\eta)$ as in Eq.~(\ref{eqn:xyTse}), we may see that:
\begin{equation}
\begin{aligned}
dx&=(\partial_\sigma x)d\sigma+(\partial_\eta x)d\eta=\left(\cos\varphi-\sigma\sin\varphi\partial_\sigma\varphi
-\eta\cos\varphi\partial_\sigma\varphi\right)d\sigma+(-\sin\varphi)d\eta,
\\
dy&=(\partial_\sigma y)d\sigma+(\partial_\eta y)d\eta=\left(\sin\varphi+\sigma\cos\varphi\partial_\sigma\varphi
-\eta\sin\varphi\partial_\sigma\varphi\right)d\sigma+(\cos\varphi)d\eta.
\end{aligned}
\end{equation}
Thus, we see that:
\begin{equation}
(ds)^2=-(cdt)^2+\left[\left(\cos\varphi-\sigma\sin\varphi\partial_\sigma\varphi
-\eta\cos\varphi\partial_\sigma\varphi\right)d\sigma-(\sin\varphi)d\eta\right]^2
+\left[\left(\sin\varphi+\sigma\cos\varphi\partial_\sigma\varphi
-\eta\sin\varphi\partial_\sigma\varphi\right)d\sigma+(\cos\varphi)d\eta\right]^2+(dz)^2.
\end{equation}
This may be seen to simplify to:
\begin{equation}
(ds)^2=-(cdt)^2+\left[(1-\eta\partial_\sigma\varphi)^2
+\sigma^2(\partial_\sigma\varphi)^2\right](d\sigma)^2+(d\eta)^2+(dz)^2,
\end{equation}
and in doing so observe that $(\sigma,\eta)$ are indeed orthogonal to one another as they ought to be. Thus, the metric of the curved surface is $g_{\mu\nu}={\rm diag}(-1,h_1^2,1,1)$ where:
\begin{equation}
h_1^2=(1-\eta\partial_\sigma\varphi)^2
+\sigma^2(\partial_\sigma\varphi)^2=(1-\eta R^{-1})^2+(\sigma R^{-1})^2,
\end{equation}
and $R=(\partial_\sigma\varphi)^{-1}$ is the local radius of curvature of the surface, which may be seen to follow from the definition of the mean curvature of the surface $R^{-1}\equiv2\kappa=-\bm{\nabla}\cdot\hat{\bm{n}}$ where $\hat{\bm{n}}$ is the unit normal of the surface. In the present case, this is none another than $\bm{e}_\eta=-\sin\varphi\bm{e}_x+\cos\varphi\bm{e}_y$ and so:
\begin{equation}
-\bm{\nabla}\cdot\bm{e}_\eta=(\partial_x\varphi)\cos\varphi+(\partial_y\varphi)\sin\varphi.
\end{equation}
Through the chain rule: $\partial_{x,y}\varphi=(\partial_\sigma\varphi)(\partial_{x,y}\sigma)$ and from the above matrix equation for $x,y$: $\partial_x\sigma=\cos\varphi$ and $\partial_y\sigma=\sin\varphi$. Thus:
\begin{equation}
-\bm{\nabla}\cdot\bm{e}_\eta=\partial_\sigma\varphi\implies R=(\partial_\sigma\varphi)^{-1}.
\end{equation} 
Finally, the determinant is $g=|g_{\mu\nu}|=-h_1^2$ and the contravariant metric is $g^{\mu\nu}={\rm diag}(-1,h_1^{-2},1,1)$.

Now, taking the equations as in Eq.~(\ref{eqn:covEM}) with this metric and:
\begin{equation}
F_{\mu\nu}=
\begin{pmatrix}
0&-E_\sigma/c&-E_\eta/c&-E_z/c
\\
E_\sigma/c&0&-B_z&B_\eta
\\
E_\eta/c&B_z&0&-B_\sigma
\\
E_z/c&-B_\eta&B_\sigma&0
\end{pmatrix},\quad
\tilde{F}_{\mu\nu}=
\begin{pmatrix}
0&-\epsilon E_\sigma/c&-\epsilon E_\eta/c&-\epsilon E_z/c
\\
\epsilon E_\sigma/c&0&-B_z&B_\eta
\\
\epsilon E_\eta/c&B_z&0&-B_\sigma
\\
\epsilon E_z/c&-B_\eta&B_\sigma&0
\end{pmatrix},
\end{equation}
(since the covariant formulation allows for the simple renaming $(x,y,z)\rightarrow(\sigma,\eta,z)$), where $\epsilon$ is the dielectric function that is free to vary in space, it may be shown through unilluminating algebra that:
\begin{align}
{\bm \nabla}\cdot(\epsilon\bm{{\cal E}})=0,&\quad{\bm \nabla}\cdot{\bm B}=0,\label{eqn:SL}
\\
\partial_t(\epsilon\bm{{\cal E}})=c^2\bm{\nabla}\times\bm{{\cal B}},&\quad \partial_t{\bm B}=-{\bm \nabla}\times{\bm E},\label{eqn:S}
\end{align}
where:
\begin{align}
\bm{\nabla}=\bm{e}_\sigma\partial_\sigma+\bm{e}_\eta\partial_\eta
+\bm{e}_z\partial_z,
\quad
\bm{{\cal E}}=\bm{e}_\sigma E_\sigma h_1^{-1}+\bm{e}_\eta E_\eta h_1+\bm{e}_zE_zh_1,
\quad
\bm{{\cal B}}=\bm{e}_\sigma B_\sigma h_1+\bm{e}_\eta B_\eta h_1^{-1}+\bm{e}_zB_zh_1^{-1}.
\end{align}

Gauss' laws in Eq.~(\ref{eqn:SL}) do not help to determine the SPP fields from the incident fields. Rather, we consider Amp{\`e}re's and Maxwell's equations Eq.~(\ref{eqn:S}). Evaluating the curls, rearranging the resultant expressions, assuming invariance in time, {\it i.e.} $\{\bm{{\cal E}},\bm{{\cal B}}\}\propto e^{i\omega t}$, and measuring all the length scales within the problem in units of $\lambdabar=\lambda/(2\pi)$ yields:
\begin{equation}
{\cal L}_{\rm TM}\bm{u}=\bm{g}_{\rm TM},\quad{\cal L}_{\rm TE}\bm{v}=\bm{g}_{\rm TE},
\end{equation}
where:
\begin{align}
{\cal L}_{\rm TM}=
\begin{pmatrix}
1&i\partial_z&-i\partial_\eta
\\
-i\partial_z&-\epsilon&0
\\
i\partial_\eta&0&-\epsilon
\end{pmatrix},\quad\bm{u}=
\begin{pmatrix}
cB_\sigma
\\
E_\eta
\\
E_z
\end{pmatrix},
&\quad
\bm{g}_{\rm TM}=
\begin{pmatrix}
0
\\
-i(\partial_\sigma v_3)h_1^{-1}
\\
i(\partial_\sigma v_2)h_1^{-1}+iu_1(Rh_1)^{-1}
\end{pmatrix},
\\
{\cal L}_{\rm TE}=
\begin{pmatrix}
-\epsilon&i\partial_z&-i\partial_\eta
\\
-i\partial_z&1&0
\\
i\partial_\eta&0&1
\end{pmatrix},\quad\bm{v}=
\begin{pmatrix}
E_\sigma
\\
cB_\eta
\\
cB_z
\end{pmatrix},
&\quad
\bm{g}_{\rm TE}=
\begin{pmatrix}
iu_3(Rh_1)^{-1}
\\
-i\partial_\sigma u_3
\\
i\partial_\sigma u_2
\end{pmatrix}.
\end{align}

At this point, the asymptotic expansion is implemented. Assuming that the surface is sufficiently smooth based upon the assumption that $R\gg\lambdabar$, we make an expansion in a smallness parameter $\alpha$ such that $R$ goes as $\alpha^2$.

Since $R=(\partial_\sigma\varphi)^{-1}$, $\varphi$ is linear in $\sigma$. Thus, we expand $\sigma$ in $\alpha$ simply as: $\sigma=\alpha\sigma_1$. Then, it is clear that $R\sim\alpha^{-2}$ since we also now have that $\partial_\sigma=\alpha\partial_{\sigma_1}$. The fields are expanded in a like manner as $\bm{u}=\bm{u}^{(0)}+\alpha\bm{u}^{(1)}+\alpha^2\bm{u}^{(2)}+\cdots$, $\bm{v}=\bm{v}^{(0)}+\alpha\bm{v}^{(1)}+\alpha^2\bm{v}^{(2)}+\cdots$, however their amplitudes are also allowed to retain corrections in $\alpha$ to avoid secular growing terms within the analysis. As a result, we write:
\begin{equation}
\bm{u}^{(0)}=F
\begin{pmatrix}
a
\\
b
\\
c
\end{pmatrix}e^{ipz}e^{iq_j\eta},\quad
\bm{v}^{(1)}=G
\begin{pmatrix}
d
\\
e
\\
f
\end{pmatrix}e^{ipz}e^{iq_j\eta},\quad
\bm{u}^{(2)}=H
\begin{pmatrix}
r
\\
s
\\
t
\end{pmatrix}e^{ipz}e^{iq_j\eta},
\end{equation}
where $Z(\sigma_1,z)=Z_0(\sigma_1)+\alpha Z_1(\sigma_1,z)+\alpha^2Z_2(\sigma_1,z)+\cdots$ with $Z\in\{F,G,H\}$ are allowed to vary with $z$ and $\sigma_1$ but we also have that $q_j=q_j^{(0)}+\alpha q_j^{(1)}+\alpha^2q_j^{(2)}+{\cal O}(\alpha^3)$ since, at all orders, we must satisfy the interface boundary condition, $\hat{\bm{n}}\cross(\bm{E}_2-\bm{E}_1)=\bm{0}$, at $\eta=0$ whereby the dielectric jumps between $\epsilon_1$ and $\epsilon_2$.

Taking the differential equations up to and including $\alpha^2$ we see that for $\bm{u}$:
\begin{equation}
\begin{pmatrix}
1&i\partial_z&-i\partial_\eta
\\
-i\partial_z&-\epsilon_j&0
\\
i\partial_\eta&0&-\epsilon_j
\end{pmatrix}
\begin{pmatrix}
Fa+Hr
\\
Fb+Hs
\\
Fc+Ht
\end{pmatrix}e^{i(pz+q_j\eta)}\\=
\begin{pmatrix}
0
\\
-i\alpha\partial_{\sigma_1}\left(\alpha Gfe^{i(pz+q_j\eta)}\right)h_1^{-1}
\\
i\alpha\partial_{\sigma_1}\left(\alpha Gee^{i(pz+q_j\eta)}\right)h_1^{-1}
+i\alpha^2Fae^{i(pz+q_j\eta)}(Rh_1)^{-1}
\end{pmatrix},
\end{equation}
whilst for $\bm{v}$ we have that:
\begin{equation}
\begin{pmatrix}
-\epsilon_j&i\partial_z&-i\partial_\eta
\\
-i\partial_z&1&0
\\
i\partial_\eta&0&1
\end{pmatrix}\alpha G
\begin{pmatrix}
d
\\
e
\\
f
\end{pmatrix}e^{i(pz+q_j\eta)}
\\=
\begin{pmatrix}
0
\\
-i\alpha\partial_{\sigma_1}\left(Fce^{i(pz+q_j\eta)}\right)h_1^{-1}
\\
i\alpha\partial_{\sigma_1}\left(Fbe^{i(pz+q_j\eta)}\right)h_1^{-1}+i\alpha^2Gde^{i(pz+q_j\eta)}(Rh_1)^{-1}
\end{pmatrix},
\end{equation}
where the fact that $R=[\alpha^2\partial_{\sigma_1}\varphi(\sigma_1)]^{-1}$ has been used.

Thus at order $\alpha^0$, it is obvious to see that:
\begin{equation}
\begin{pmatrix}
1&i\partial_z&-i\partial_\eta
\\
-i\partial_z&-\epsilon_j&0
\\
i\partial_\eta&0&-\epsilon_j
\end{pmatrix}F_0
\begin{pmatrix}
a
\\
b
\\
c
\end{pmatrix}e^{i(pz+q_j^{(0)}\eta)}
=\bm{0},
\end{equation}
in which case we arrive at:
\begin{equation}
\begin{aligned}
a-pb+q_j^{(0)}c&=0,
\\
pa-\epsilon_j b&=0,
\\
-q_j^{(0)}a-\epsilon_j c&=0,
\end{aligned}
\end{equation}
The solution to which is simply $a=1$, $b=p\epsilon_j^{-1}$ and $c=-q^{(0)}_j\epsilon_j^{-1}$. Thus, we see that:
\begin{equation}
\bm{u}^{(0)}=(F_0+\alpha^2F_2)
\begin{pmatrix}
1
\\
p\epsilon_j^{-1}
\\
-q_j^{(0)}\epsilon_j^{-1}
\end{pmatrix}e^{i(pz+q_j\eta)}
\end{equation}
where the correct signs of $q_j$ must be chosen to ensure the correct behaviour of the function at infinity when in either of $\eta>0$, wherein $j=1$, or $\eta<0$, wherein $j=2$. We find $p$ and $q$ by making use of the interface condition upon the electromagnetic field at the $\eta=0$ to see that $q^{(0)}_1\epsilon_2=q^{(0)}_2\epsilon_1$. Thus, we find $p$ and $q_j^{(0)}$ through:
\begin{equation}
\epsilon_j=p^2+\left(q_j^{(0)}\right)^2,\quad q_1^{(0)}\epsilon_2=q_2^{(0)}\epsilon_1,
\end{equation}
as:
\begin{equation}
p=\pm\sqrt{\frac{\epsilon_1\epsilon_2}{\epsilon_1+\epsilon_2}},\quad q_j^{(0)}=\pm\sqrt{\frac{\epsilon_j^2}{\epsilon_1+\epsilon_2}}.
\end{equation}

Before continuing, the solvability condition for a differential equation must be introduced. This is the universal statement that, given an entirely general differential operator ${\cal L}_{\bm{x}}$ acting over a vector space $\bm{x}$ that describes the following differential equation:
\begin{equation}
{\cal L}_{\bm{x}}\bm{w}^{\rm R}(\bm{x})=\bm{Y}(\bm{x}),
\end{equation}
a solution may exist if and only if:
\begin{equation}\label{eqn:solvC}
\langle \bm{w}^{\rm L}_0,{\bm Y}\rangle=0,
\end{equation}
where $\bm{w}^{\rm L}_0(\bm{x})$ is the left-eigenvector that solves the homogeneous differential equation:
\begin{equation}\label{eqn:solvChom}
{\cal L}_{\bm x}^{\rm T}\left[\bm{w}^{\rm L}_0(\bm{x})\right]^{\rm T}=\bm{0},
\end{equation}
and the inner product signifies to take an integral over the complete vector space $\bm{x}$.

At first order in $\alpha$, we see that we have:
\begin{equation}
\begin{pmatrix}
-\epsilon_j&i\partial_z&-i\partial_\eta
\\
-i\partial_z&1&0
\\
i\partial_\eta&0&1
\end{pmatrix}\alpha(G_0+\alpha G_1)
\begin{pmatrix}
d
\\
e
\\
f
\end{pmatrix}e^{i(pz+q^{(0)}_j\eta+\alpha q^{(1)}_j\eta)}=
\begin{pmatrix}
0
\\
-i[\alpha\partial_{\sigma_1}\left(-(F_0+\alpha F_1)q^{(0)}_j\epsilon_j^{-1}e^{i(pz+q^{(0)}_j\eta+\alpha q^{(1)}_j\eta)}\right)]
\\
i[\alpha\partial_{\sigma_1}\left((F_0+\alpha F_1)p\epsilon_j^{-1}e^{i(pz+q^{(0)}_j\eta+\alpha q^{(1)}_j\eta)}\right)].
\end{pmatrix},
\end{equation}
which reduces to:
\begin{equation}
(G_0+\alpha G_1)
\begin{pmatrix}
-\alpha\epsilon_j&-\alpha p+i\alpha^2(\partial_zG_1)(G_0+\alpha G_1)^{-1}&\alpha q_j^{(0)}+\alpha^2q_j^{(1)}
\\
\alpha p-i\alpha^2(\partial_zG_1)(G_0+\alpha G_1)^{-1}&\alpha&0
\\
-\alpha q_j^{(0)}-\alpha^2q_j^{(1)}&0&\alpha
\end{pmatrix}
\begin{pmatrix}
d
\\
e
\\
f
\end{pmatrix}=
\begin{pmatrix}
0
\\
i\partial_{\sigma_1}(\alpha F_0+\alpha^2F_1)q_j^{(0)}\epsilon_j^{-1}
\\
i\partial_{\sigma_1}(\alpha F_0+\alpha^2F_1)p\epsilon_j^{-1}
\end{pmatrix},
\end{equation}
since $h_1=1$ at order $\alpha^1$ as $h_1=1-\eta/R$ and $R\sim1/\alpha^2$. 

Now, taking this equation to separate orders in $\alpha$ we see:
\begin{equation}
G_0\begin{pmatrix}
\epsilon_j&-p&q_j^{(0)}
\\
p&1&0
\\
-q_j^{(0)}&0&1
\end{pmatrix}
\begin{pmatrix}
d
\\
e
\\
f
\end{pmatrix}=
\begin{pmatrix}
0
\\
i(\partial_{\sigma_1}F_0)q_j^{(0)}\epsilon_j^{-1}
\\
i(\partial_{\sigma_1}F_0)p\epsilon_j^{-1}
\end{pmatrix},
\end{equation}
and:
\begin{equation}
G_1\begin{pmatrix}
-\epsilon_j&-p+i(\partial_zG_1)G_1^{-1}&q_j^{(0)}+q_j^{(1)}G_0G_1^{-1}
\\
p-i(\partial_zG_1)G_1^{-1}&1&0
\\
-q_j^{(0)}-q_j^{(1)}G_0G_1^{-1}&0&1
\end{pmatrix}
\begin{pmatrix}
d
\\
e
\\
f
\end{pmatrix}=
\begin{pmatrix}
0
\\
i(\partial_{\sigma_1}F_1)q_j^{(0)}\epsilon_j^{-1}
\\
i(\partial_{\sigma_1}F_1)p\epsilon_j^{-1}
\end{pmatrix}.
\end{equation}
Taking the solvability condition as introduced in the above, the left-eigenvector that solves the homogeneous equation involving $G_1$ has components:
\begin{equation}
d=-1,\quad e=-i(\partial_zG_1)G_1^{-1}+p,\quad f=-q_j^{(0)}-q_j^{(1)}G_0G_1^{-1}.
\end{equation}
Thus, the solvability condition yields:
\begin{equation}
0=\int d\sigma_1d\eta dz\left[0+\left(\frac{-i(\partial_zG_1)}{G_1}+p\right)\frac{i(\partial_{\sigma_1}F_1)q_j^{(0)}}{G_1\epsilon_j}-\left(q_j^{(0)}+\frac{q_j^{(1)}G_0}{G_1}\right)\frac{i(\partial_{\sigma_1}F_1)p}{G_1\epsilon_j}\right],
\end{equation}
which reduces to:
\begin{equation}
\int d\sigma_1d\eta dz\left[(\partial_zG_1)q_j^{(0)}-iq_j^{(1)}G_0p\right]\frac{(\partial_{\sigma_1}F_1)}{G_1^2\epsilon_j}=0.
\end{equation}
In addition:
\begin{equation}
\epsilon_j-\left(p+i(\partial_zG_1)G_1^{-1}\right)^2-\left(q_j^{(0)}+q_j^{(1)}G_0G_1^{-1}\right)^2=0,
\end{equation}
from the first line of the above matrix equation involving $G_1$. Thus:
\begin{equation}
-2ip(\partial_zG_1)G_1^{-1}-2q_j^{(0)}q_j^{(1)}G_0G_1^{-1}+(\partial_zG_1)^2G_1^{-2}
-\left(q_j^{(1)}\right)^2G_0^2G_1^{-2}=0.
\end{equation}
From the solvability condition, $\partial_zG_1=iq_j^{(1)}G_0p/q_j^{(0)}$ and so:
\begin{equation}
2p^2q_j^{(1)}G_0G_1-2\left(q_j^{(0)}\right)^2q_j^{(1)}G_0G_1
-\left(q_j^{(1)}\right)^2G_0^2p^2/q_j^{(0)}
-\left(q_j^{(1)}\right)^2G_0^2=0.
\end{equation}
Now, since this equation is multiplied overall by $q_j^{(1)}$ and all other terms are known to be non-zero it must be that $q_j^{(1)}=0$ from which $\partial_zG_1=0$ by the solvability condition.


So, now at order $\alpha^1$:
\begin{equation}
\begin{pmatrix}
\epsilon_j&p&-q_j^{(0)}
\\
-p&-1&0
\\
q_j^{(0)}&0&-1
\end{pmatrix}
\begin{pmatrix}
d
\\
e
\\
f
\end{pmatrix}=-i\frac{\partial_{\sigma_1} F_0}{G_0}
\begin{pmatrix}
0
\\
q_j^{(0)}\epsilon_j^{-1}
\\
p\epsilon_j^{-1}
\end{pmatrix},
\end{equation}
and at order $\alpha^2$:
\begin{equation}
\begin{pmatrix}
\epsilon_j&p&-q_j^{(0)}
\\
-p&-1&0
\\
q_j^{(0)}&0&-1
\end{pmatrix}
\begin{pmatrix}
d
\\
e
\\
f
\end{pmatrix}=-i\frac{\partial_{\sigma_1} F_1}{G_1}
\begin{pmatrix}
0
\\
q_j^{(0)}\epsilon_j^{-1}
\\
p\epsilon_j^{-1}
\end{pmatrix},
\end{equation}
which have the solutions $d=i(\partial_{\sigma_1} F_k)q_j^{(0)}(G_kp\epsilon_j)^{-1}$, $e=0$, $f=i(\partial_{\sigma_1} F_k)(G_kp)^{-1}$, where the choice $e=0$ is such that $\hat{\bm{n}}\cdot(\bm{B}_2-\bm{B}_1)=0$ holds, and so the first correction to $\bm{v}$ is:
\begin{equation}
\bm{v}^{(1)}=\left[i(\partial_{\sigma_1}F_0)+\alpha i(\partial_{\sigma_1}F_1)\right]
\begin{pmatrix}
q^{(0)}_j(p\epsilon_j)^{-1}
\\
0
\\
p^{-1}
\end{pmatrix}e^{i(pz+q_j\eta)}=i(\partial_{\sigma_1}F)
\begin{pmatrix}
q^{(0)}_j(p\epsilon_j)^{-1}
\\
0
\\
p^{-1}
\end{pmatrix}e^{i(pz+q_j\eta)}.
\end{equation}

Note that the determinant of the matrix on the left-hand side in the above is equal to zero. Thus, the solution is not unique; we could equally well have chosen any of the other components to be zero. However, we would have found a consistent result in either case. This may be seen by considering Gauss' law for magnetism $\bm{\nabla}\cdot\bm{B}=0$ to find that, given the above expansion, $\partial_\eta B_\eta=0$ showing that our choice was consistent.

Now, taking the differential equation for $\bm{u}$ to order $\alpha^2$:
\begin{multline}
F_0
\begin{pmatrix}
1&-p&q_j^{(0)}+\alpha^2q_j^{(2)}
\\
p&-\epsilon_j&0
\\
-q_j^{(0)}-\alpha^2q_j^{(2)}&0&-\epsilon_j
\end{pmatrix}
\begin{pmatrix}
1
\\
p\epsilon_j^{-1}
\\
-q_j^{(0)}\epsilon_j^{-1}
\end{pmatrix}
+\alpha F_1
\begin{pmatrix}
1&-p+i(\partial_zF_1)F_1^{-1}&q_j^{(0)}+\alpha^2q_j^{(2)}
\\
p-i(\partial_zF_1)F_1^{-1}&-\epsilon_j&0
\\
-q_j^{(0)}-\alpha^2q_j^{(2)}&0&-\epsilon_j
\end{pmatrix}
\begin{pmatrix}
1
\\
p\epsilon_j^{-1}
\\
-q_j^{(0)}\epsilon_j^{-1}
\end{pmatrix}
\\
+\alpha^2F_2
\begin{pmatrix}
1&-p+i(\partial_zF_2)F_2^{-1}&q_j^{(0)}+\alpha^2q_j^{(2)}
\\
p-i(\partial_zF_2)F_2^{-1}&-\epsilon_j&0
\\
-q_j^{(0)}-\alpha^2q_j^{(2)}&0&-\epsilon_j
\end{pmatrix}
\begin{pmatrix}
1
\\
p\epsilon_j^{-1}
\\
-q_j^{(0)}\epsilon_j^{-1}
\end{pmatrix}
\\+\alpha^2H_0
\begin{pmatrix}
1&-p&q^{(0)}+\alpha^2q_j^{(2)}
\\
p&-\epsilon_j&0
\\
-q_j^{(0)}-\alpha^2q_j^{(2)}&0&-\epsilon_j
\end{pmatrix}
\begin{pmatrix}
r
\\
s
\\
t
\end{pmatrix}
=\alpha^2
\begin{pmatrix}
0
\\
(\partial_{\sigma_1\sigma_1}F_0)p^{-1}
\\
iF_0R^{-1}
\end{pmatrix},
\end{multline}
since $\alpha^2h_1^{-1}=1$ when the resultant $\alpha^4$ term is ignored and $F_1=0$ may be seen since those are the only terms proportional to $\alpha$. So to all orders up to and including $\alpha^2$ (since the $\alpha^0$ and $\alpha^1$ terms all identically cancel to zero) we see that:
\begin{equation}
H_0
\begin{pmatrix}
1&-p&q_j^{(0)}
\\
p&-\epsilon_j&0
\\
-q_j^{(0)}&0&-\epsilon_j
\end{pmatrix}
\begin{pmatrix}
r
\\
s
\\
t
\end{pmatrix}=
\begin{pmatrix}
-i(\partial_zF_2)p\epsilon_j^{-1}+F_0q_j^{(0)}q_j^{(2)}\epsilon_j^{-1}
\\
i(\partial_zF_2)+(\partial_{\sigma_1\sigma_1}F_0)p^{-1}
\\
F_0q_j^{(2)}+iF_0R^{-1}
\end{pmatrix}=
\begin{pmatrix}
A_j
\\
B_j
\\
C_j
\end{pmatrix}
\end{equation}
Likewise to the above, the matrix on the left-hand side has zero determinant and so we must choose a component to be vanishing as our gauge choice. We choose $r$ to simplify the analysis and it may be shown that this is consistent through Gauss' law.

Then we find the solutions for the fields as:
\begin{equation}
r=0,\quad s=-\frac{B_j}{H_0\epsilon_j},\quad t=-\frac{C_j}{H_0\epsilon_j},
\end{equation}
and thus:
\begin{equation}
\bm{u}^{(2)}=
\begin{pmatrix}
0
\\
-B_j\epsilon_j^{-1}
\\
-C_j\epsilon_j^{-1}
\end{pmatrix}e^{i(pz+q_j\eta)}.
\end{equation}
Taking Gauss' law as we have: $\bm{\nabla}\cdot(\epsilon\bm{{\cal E}})=0$, it may be seen that $-(\partial_zF_0)(q_j^{(2)}+iR^{-1})=0$, which holds since $F_0$ does not depend on $z$.

To find $q_j^{(2)}$ we make use of the interface condition $\hat{\bm{n}}\cross(\bm{E}_2-\bm{E}_1)=\bm{0}$ at $\eta=0$ as:
\begin{equation}
\bm{e}_\eta\cross\left[\bm{e}_\sigma(E_{2\sigma}-E_{1\sigma})
+\bm{e}_\eta(E_{2\eta}-E_{1\eta})+\bm{e}_z(E_{2z}-E_{1z})\right]=\bm{e}_z(E_{1\sigma}-E_{2\sigma})+\bm{e}_\sigma(E_{2z}-E_{1z})=\bm{0},
\end{equation}
which, with our fields, becomes:
\begin{equation}
\left[v_1^{(1)}\right]_1=\left[v_1^{(1)}\right]_2,\quad\left[u_3^{(0)}\right]_1
+\alpha^2\left[u_3^{(2)}\right]_1=\left[u_3^{(0)}\right]_2+\alpha^2\left[u_3^{(2)}\right]_2.
\end{equation}
Since we already have that $q_1^{(0)}\epsilon_2=q_2^{(0)}\epsilon_1$ then the terms proportional to $\alpha^0$ and $\alpha^1$ cancel. Thus, we are left with (when $\eta=0$):
\begin{equation}
i\alpha(\partial_{\sigma_1}F_0)\frac{q_1^{(0)}}{p\epsilon_1}=i\alpha(\partial_{\sigma_1}F_0)\frac{q_2^{(0)}}{p\epsilon_2},\quad-\frac{q_1^{(0)}}{\epsilon_1}+\alpha^2\frac{C_1}{\epsilon_1}=-\frac{q_2^{(0)}}{\epsilon_2}+\alpha^2\frac{C_2}{\epsilon_2},
\end{equation}
which reduces to:
\begin{equation}
0=0,\quad\left(q_1^{(2)}+iR^{-1}\right)\epsilon_2
=\left(q_2^{(2)}+iR^{-1}\right)\epsilon_1,
\end{equation}
If we take our results for the components of the $\bm{u}^{(2)}$ field: $r$, $s$ and $t$, and substitute them into the top line of their matrix equation then we see that:
\begin{equation}
r-ps+q_j^{(0)}t=A_j,
\end{equation}
reduces to the following:
\begin{equation}
2p\frac{\partial F_2}{\partial z}-i\frac{\partial^2F_0}{\partial\sigma_1^2}+q_j^{(0)}F_0\left(2iq_j^{(2)}-R^{-1}\right)=0.
\end{equation}
This equation must be satisfied when both $\eta>0$ and $\eta<0$ thus it follows that:
\begin{equation}
q_1^{(0)}\left(2q_1^{(2)}+iR^{-1}\right)=q_2^{(0)}\left(2q_2^{(2)}+iR^{-1}\right),
\end{equation}
thus we have two simultaneous equations for $q_1^{(2)}$ and $q_2^{(2)}$, which we may solve to find that:
\begin{equation}
q_j^{(2)}=i\left[\frac{\left(q_1^{(0)}+q_2^{(0)}\right)\epsilon_j-2q_j^{(0)}\epsilon_j}{2R\left(q_1^{(0)}\epsilon_1-q_2^{(0)}\epsilon_2\right)}\right].
\end{equation}
or, in a simpler form:
\begin{equation}
q_1^{(2)}=\frac{-i(\epsilon_1+2\epsilon_2)}{2R(\epsilon_1+\epsilon_2)},\quad q_2^{(2)}=\frac{-i(2\epsilon_1+\epsilon_2)}{2R(\epsilon_1+\epsilon_2)}.
\end{equation}

We now make use of the solvability condition again that must apply to this differential equation. Firstly, we find the left-eigenvector of:
\begin{equation}
\bm{w}_0^{\rm L}(\bm{x})
\begin{pmatrix}
1&-p&q_j^{(0)}
\\
p&-\epsilon_j&0
\\
-q_j^{(0)}&0&-\epsilon_j
\end{pmatrix}=\bm{0},
\end{equation}
where $\bm{w}_0^{\rm L}(\bm{x})=(\tilde{r},\tilde{s},\tilde{t})$, as the following simultaneous equations:
\begin{equation}
\begin{aligned}
\tilde{r}+p\tilde{s}-q_j^{(0)}\tilde{t}&=0,
\\
-p\tilde{r}-\epsilon_j\tilde{s}&=0,
\\
q_j^{(0)}\tilde{r}-\epsilon_j\tilde{t}&=0,
\end{aligned}
\end{equation}
whose solution is: $\tilde{r}=1$, $\tilde{s}=-p\epsilon_j^{-1}$, $\tilde{t}=q_j^{(0)}\epsilon_j^{-1}$. Thus, we now make use of the solvability condition as:
\begin{multline}
0=\int d\sigma_1d\eta dz
\begin{pmatrix}
\tilde{r}&\tilde{s}&\tilde{t}
\end{pmatrix}
\begin{pmatrix}
A_j
\\
B_j
\\
C_j
\end{pmatrix}=
\int d\sigma_1 d\eta dz
\begin{pmatrix}
1&-p\epsilon_j^{-1}&q_j^{(0)}\epsilon_j^{-1}
\end{pmatrix}
\begin{pmatrix}
-i(\partial_zF_2)p\epsilon_j^{-1}+F_0q_j^{(0)}q_j^{(2)}\epsilon_j^{-1}
\\
i(\partial_zF_2)+(\partial_{\sigma_1\sigma_1}F_0)p^{-1}
\\
F_0q_j^{(2)}+iF_0R^{-1}
\end{pmatrix}\\=
\int d\sigma_1d\eta dz\Big[-2i(\partial_z F_2)p\epsilon_j^{-1}-(\partial_{\sigma_1\sigma_1}F_0)\epsilon_j^{-1}
+2F_0q_j^{(0)}q_j^{(2)}\epsilon_j^{-1}+iF_0q_j^{(0)}(R\epsilon_j)^{-1}\Big].
\end{multline}
Thus, we find the following differential equation linking the amplitudes $F_0,F_2$:
\begin{equation}
2p\frac{\partial F_2}{\partial z}-i\frac{\partial^2F_0}{\partial\sigma_1^2}+q_j^{(0)}F_0\left(2iq_j^{(2)}-R^{-1}\right)=0,
\end{equation}
which is the self-same differential equation as found from $r-ps+q_j^{(0)}t=A_j$.
Substituting our expression for $q_j^{(2)}$ from the above into this equation yields:
\begin{equation}
2p\frac{\partial F_2}{\partial z}-i\frac{\partial^2F_0}{\partial\sigma_1^2}+\frac{p^2q_j^{(0)}}{R\epsilon_j}F_0=0.
\end{equation}

At this point we terminate the asymptotic expansion since we have arrived at the order that corresponds to the radius of curvature $R^{-1}\sim \alpha^2$ and so drop the expansion indices. Furthermore, we identify that $p=n_e$ is the effective index of the SPP and that $q_j=i\gamma_j$ since $\epsilon_j^2/(\epsilon_1+\epsilon_2)$ will always be a negative number at the allowed frequencies of SPP resonance. Finally, we reintroduce the dimensionality within the problem in terms of $\lambdabar$ and so find:
\begin{equation}
i\lambdabar\frac{\partial F}{\partial z}=-\frac{\lambdabar^2}{2n_e}\frac{\partial^2F}{\partial\sigma^2}+V(\sigma)F,
\end{equation}
where:
\begin{equation}
V(\sigma)=\frac{\lambdabar n_e}{2R(\sigma)}\sqrt{\frac{-1}{\epsilon_1+\epsilon_2}},
\end{equation}
and $R$, being proportional $\partial_\sigma\varphi(\sigma)$, varies step-wise along the surface even when the radii of curvature of the peaks and troughs varies in magnitude so long as the aperture angle is the same for all.

Then, if we take the SPP that `forward' scatterers with a positive effective index of  $p=+n_e$, and assume translational invariance in $z$ such that we may say $F(\sigma,z)=F(\sigma)e^{i\Delta n z/\lambdabar}$, then we see:
\begin{equation}
-\Delta n F=-\frac{\lambdabar^2}{2n_e}\frac{\partial^2F}{\partial\sigma^2}+V(\sigma)F.
\end{equation}
Conversely, if we choose the SPP that `backward' scatterers with negative index $p=-n_e$  then we say that $F(\sigma,z)=F(\sigma)e^{-i\Delta n z/\lambdabar}$ and so:
\begin{equation}
\Delta n F=\frac{\lambdabar^2}{2n_e}\frac{\partial^2F}{\partial\sigma^2}-V(\sigma)F.
\end{equation}
In other words, both values for $p$ are described by the same Schr{\"o}dinger-like equation for negative eigenvalues. The corrected index is then $n=n_e+\Delta n$ in either case whose sign simply indicates the propagation direction along $z$.

\end{document}